\documentclass[onecolumn,twoside]{IEEEtran}
\usepackage{mathpazo}
\usepackage{times}
\usepackage{amsmath}
\usepackage{amsfonts}
\usepackage{latexsym}
\usepackage{amssymb}
\usepackage{mathrsfs}

\usepackage{upref}
\usepackage{theorem}
\usepackage{graphicx}
\usepackage{psfrag}
\usepackage{cite}

\usepackage{color}

\theoremstyle{plain}
\theorembodyfont{\normalfont\slshape}

\newtheorem{thm}{Theorem$\!$}
\newenvironment{theorem}
{\begin{thm}\hspace*{-1ex}{\bf.}}{\end{thm}}

\newtheorem{lem}[thm]{Lemma$\!$}
\newenvironment{lemma}{\begin{lem}\hspace*{-1ex}{\bf.}}{\end{lem}}

\newtheorem{alg}[thm]{Algorithm$\!$}
\newenvironment{algorithm}{\begin{alg}\hspace*{-1ex}{\bf.}}{\end{alg}}

\newtheorem{prop}[thm]{Proposition$\!$}

\newtheorem{cor}[thm]{Corollary$\!$}

\newtheorem{defn}[thm]{Definition$\!$}
\newenvironment{definition}{\begin{defn}\hspace*{-1ex}{\bf.}}{\end{defn}}

\newtheorem{xmpl}[thm]{Example$\!$}
\newenvironment{example}{\begin{xmpl}\hspace*{-1ex}{\bf.}}{\hfill$\Box$\end{xmpl}}

\newtheorem{cnstr}{Construction$\!$}

\newcounter{enumrom}
\renewcommand{\theenumrom}{(\roman{enumrom})}

\makeatletter
\renewcommand{\@endtheorem}{\endtrivlist}
\makeatother

\makeatletter
\renewcommand{\thefigure}{{\@arabic\c@figure}}
\renewcommand{\fnum@figure}{{\bf Figure\,\thefigure}}
\makeatother

\newcommand{\cC}{\mathcal{C}}

\newcommand{\cH}{\mathcal{H}}

\newcommand{\cO}{\mathcal{O}}

\newcommand{\pf}{{\bf Proof: }}

\newcommand{\uw}{\mbox{$\underline{w}$}}
\newcommand{\uv}{\mbox{$\underline{v}$}}

\newcommand{\be}[1]{\begin{equation}\label{#1}}
\newcommand{\ee}{\end{equation}}

\renewcommand{\leq}{\leqslant}

\renewcommand{\geq}{\geqslant}

\newcommand{\Cref}[1]{Co\-ro\-lla\-ry\,\ref{#1}}

\newcommand{\C}{\mbox{${\cal C}$}}

\newcommand{\qed}{\hfill$\Box$\\[1ex]}
\newcommand{\hs}{\mbox{$\hat{s}$}}
\newcommand{\uc}{\mbox{$\underline{c}$}}
\newcommand{\uu}{\mbox{$\underline{u}$}}
\newcommand{\al}{\alpha}

\newcommand{\xor}{\oplus}
\newcommand{\la}{\leftarrow}

\newcommand{\eq}{\mbox{$\,=\,$}}
\newcommand{\ga}{\mbox{$\gamma$}}

\outer\def\proclaim #1. #2\par{\medbreak
 \noindent{\bf#1.\enspace}{\sl#2\par}%
 \ifdim\lastskip<\medskipamount \removelastskip\penalty55\medskip\fi}

\begin{document}


\title{\Huge\bf Extended Product and Integrated Interleaved Codes}

\author{\large
Mario~Blaum and Steven~Hetzler\\
IBM Research Division\\
Almaden Research Center\\
 San Jose, CA 95120, USA \\
Mario.Blaum@ibm.com, hetzler@us.ibm.com}
\maketitle

\begin{abstract}
A new class of codes, Extended Product (EPC) Codes, consisting of a
product code with a number of extra parities added, is presented and
applications for erasure decoding are discussed. An upper
bound on the minimum distance of EPC codes is given, as well as
constructions meeting the bound for some relevant cases. A special case of
EPC codes, Extended Integrated Interleaved (EII) codes, which naturally unify
Integrated Interleaved (II) codes and product codes, is defined and
studied in detail. It is shown that
EII codes often improve the minimum distance of II codes with the same
rate, and they enhance the decoding algorithm by allowing decoding on
columns as well as on rows. It is also shown that EII codes allow for
encoding II codes with an uniform distribution of the parity symbols.
\end{abstract}


\section{Introduction}
\label{Introduction}

There has been considerable research in recent literature on
codes with local and global properties for erasure correction (see
for instance~\cite{bhh,bh,bpsy,ghsy,hcl,kna,pd,pklk,rk,rp,sa,sd,tb,wz}
and references within).
In general, data symbols are divided into sets and parity symbols
(i.e., local parities) are added to each set (often using an MDS code).
This way, when a number of erasures not exceeding the
number of parity symbols occurs in a set, such erasures are rapidly
recovered. In addition to the local parities, a number of
global parities (also called heavy parities) are added. Those
global parities involve all of the
data symbols and may include the local parity symbols as well. The
goal of the global parities is to correct situations in which the
erasure-correcting power of the local parities has been exceeded.

The interest in erasure correcting codes with local and global
properties arises mainly from
two applications. One of them is the cloud. A cloud configuration
may consist of many storage devices, of which some of them may even
be in different geographical locations, and the data is distributed
across them. If one or more of those devices fails, it is
desirable to recover its contents ``locally,'' that is, using a few
parity devices within a set of limited size in order to affect
performance as little as possible. However, the local parities may not
suffice. Extra protection is needed in case the erasure-correcting
capability of a local set is
exceeded. To address this
situation, some devices consisting of global parities are
incorporated, and when the local correction power is exceeded,
the global parity devices
are invoked and correction is attempted. If such a situation occurs,
although there will be an impact on performance,
data loss may be averted. It is expected that the cases in which
the local parity is
exceeded are relatively rare events, so the aforementioned impact on
performance does not occur frequently.
As an example of this type of application, we refer the reader to the
description of the Azure system~\cite{hsx} or to the Xorbas code
presented in~\cite{sa}.

A second application occurs in the context of Redundant Arrays of
Independent Disks (RAID) architectures~\cite{g}. In this case, a RAID
architecture protects against one or more storage device failures.
For example, RAID 5 adds one extra parity device, allowing for the
recovery of the contents of one failed device, while RAID 6 protects
against up to two device failures. In particular, if those devices
are Solid State Drives (SSDs), like flash memories, their
reliability decays with time and with the number of writes and
reads~\cite{M}.
The information in SSDs is generally divided into pages, each page
containing its own internal Error-Correction Code (ECC). It may
happen that a particular page degrades and its ECC is exceeded.
However, the user may not become aware of this situation until the page is
accessed (what is known as a silent failure). Assuming an SSD has
failed in a RAID~5 scheme, if during reconstruction a silent page
failure is encountered in one of the surviving SSDs, then data loss
will occur. A method around this situation is using RAID~6. However,
this method is costly, since it requires two whole SSDs as parity. It is more
desirable to divide the information in a RAID type of architecture
into $m\times n$ stripes: $m$ represents the size of a stripe, and
$n$ is the number of SSDs. The RAID architecture may be viewed as
consisting of a large number of stripes, each stripe encoded and
decoded independently. Certainly, codes like the ones used in cloud
applications may be used as well for RAID applications. In practice,
the choice of code depends on the
statistics of errors and on the frequency of silent page failures.
RAID systems, however, may behave differently than a cloud array of devices, in
the sense that each column represents a whole storage device. When a device
fails, then the whole column is lost, a correlation that may not
occur in cloud applications. For that reason, RAID architectures may
benefit from a special class of codes with local and global
properties, the so called Sector-Disk (SD)
codes, which take into
account such correlations~\cite{hy,ld,pb,pbh}.

From now on, we will call the entries of the codes considered in the
paper ``symbols''. Such symbols can be whole devices (for example, in
the case of cloud applications) or pages (in the case of RAID
applications for SSDs). Each symbol may be protected
by one local group, but a natural extension is to consider multiple
localities~\cite{rp,tb,zy}. Product codes~\cite{ms} represent a
special case of multiple localities: any symbol is protected by both
horizontal and vertical parities.

Product codes by themselves may also be used in RAID-type of
architectures: the horizontal parities protect a number of devices
from failure. The vertical parities allow for rapid recovery of a page
or sector within a device (a first responder type of approach).
However, if the number of silent failures exceeds the correcting
capability of the vertical code, and the horizontal code is unusable
due to device failure, data loss will occur. For that reason, it may be
convenient to incorporate a number of extra global parities to the product
code. In general, we will simply call extra parities these extra
global parities in order to avoid confusion, since in a product code
the parities on parities, by affecting all of the symbols, can also
be considered as global parities.

In effect, consider a product code consisting of $m\times
n$ arrays such that each column has $v$ parity symbols and each row
has $h$ parity symbols. If in addition to the horizontal and vertical
parities there are $g$ extra parities, we say that the code is an
Extended Product (EPC) code and we denote it by $EP(m,v;n,h;g)$.
Notice that, in particular, $EP(m,v;n,h;0)$ is a regular product
code, while $EP(m,0;n,h;g)$ is a Locally Recoverable (LRC)
code~\cite{ghjy,tb}.

Constructions of LRC codes involve different issues and tradeoffs,
like the size of the field and optimality criteria. The same is true
for EPC codes, of which, as we have seen above, LRC codes are a
special case. In particular, one goal is to keep the size of the required
finite field small, since operations over a small field have
less complexity than ones over a larger field due to the smaller look-up
tables involved. For example, Integrated Interleaved
(II) codes~\cite{hapkt,tk}
over $GF(q)$, where $q> \max\{m,n\}$, were proposed in~\cite{bh}
as LRC codes (II codes are closely related to Generalized
Concatenated Codes~\cite{bz,z}). Let us mention also the construction
in~\cite{ll} (STAIR codes), which reduces field
size when failures are correlated. Similarly, we propose a new
family of codes that
we call Extended Integrated Interleaved (EII) codes, to be defined in
Section~\ref{GP}, of which both product codes
and II codes are special cases. In earlier versions, we called such
codes Generalized Product Codes~\cite{bh2}. However, there are
several ways of generalizing product codes, and the term Generalized
Product Code usually refers to graph theoretic
constructions~\cite{gl,hp}. The new denomination avoids confusion.

As is the case with LRC codes, construction of EPC codes involves
optimality issues. For example,
LRC codes optimizing the minimum distance were presented
in~\cite{tb}. Except for special cases, II codes are
not optimal as LRC codes, but the codes in~\cite{tb} require a field
of size at least $mn$, so there is a tradeoff. The same happens with
EII codes:  except for special cases to be presented in
Section~\ref{optimality}, they do
not optimize the minimum
distance.

There are stronger criteria for optimization than the minimum
distance in LRC codes. For example, PMDS
codes~\cite{bhh,bpsy,ghjy,hy,hsx} satisfy the
Maximally Recoverable (MR) property\cite{ghjy,ghswy}. The
definition of the MR
property is extended for EPC codes in~\cite{ghswy}, but it turns out
that EPC codes with the MR property are difficult to obtain. For
example, in~\cite{ghswy} it was proven that an EPC code
$EP(n,1;n,1;1)$ (i.e., one vertical and one horizontal parity per column
and row and one extra parity) with the MR property requires a field
whose size is superlinear on $n$. We do not address EPC codes with the MR
property in this paper.

Although the constructions we present can be extended to finite fields of any
characteristic, for simplicity, we assume that they
have characteristic 2.

The paper is structured as follows: in Section~\ref{GP} we present
the definition of EII codes and give their properties, like their
erasure-correcting capability, their minimum distance and encoding and
decoding algorithms. We also show that EII codes effectively enhance
the decoding power of regular II codes, by allowing decoding on rows
as well as on columns. As another application of EII codes, we show
that II codes admit a balanced distribution of parity symbols.
In Section~\ref{optimality}, we present an upper
bound on the minimum distance of EPC codes. We show that this bound
generalizes the known bound on the minimum
distance of LRC codes.
We also present some constructions of EPC
codes optimizing the
minimum distance for $E(m,1;n,1;g)$ codes.
We end the paper by drawing some conclusions.

\section{Extended Integrated Interleaved (EII) Codes}
\label{GP}

This section is divided into subsections as follows: in Subsection~\ref{defEII} we
give the definition of EII codes and we illustrate it with several
examples. In Subsection~\ref{EDecod} we present the main (erasure) decoding 
algorithm of EII codes consisting of a triangulation process. In
Subsection~\ref{dimdist}, we give the dimension and the minimum
distance of EII codes, as well as an encoding algorithm.
In Subsection~\ref{transp} we show that the transpose arrays of the arrays in an
EII code also constitute an EII code, and this property allows for an
enhancement of the decoding algorithm, since arrays can now be
iteratively decoded on rows as well as on columns, a process that
generalizes the well known row-column iterative decoding of product
codes. As a second application of this property, we show that EII
codes allow for an uniform distribution of the parity symbols. In
Subsection~\ref{errorerasure}, we show how to extend the
erasure decoding algorithm to errors together with erasures.

\subsection{Definition of Extended Integrated Interleaved (EII) Codes}
\label{defEII}
We start by defining EII codes, which unify
product codes and II codes. II codes may be
interpreted as $m\times n$ arrays such that each row belongs in a
code $\cC_0$, and certain linear combinations of the rows belong in
nested subcodes of $\cC_0$~\cite{bh,tk,w,z2}. In addition, we assume
that each column is also in a (vertical) code, making the arrays a
subcode of a product code. We assume that the individual codes are
Reed-Solomon~\cite{ms} (RS) type of codes.
Explicitly,

\begin{definition}
\label{defGPMDS}
{\rm
Take $t+1$ integers $$0\leq u_0<u_1<\ldots <u_{t-1}<u_t\eq n$$ and let
$\uu$ be the following vector of length  
$m=s_0+s_1+\cdots +s_{t-1}+s_t$, where $s_i\geq 1$ for $0\leq
i\leq t-1$ and $s_t\geq 0$:

\begin{eqnarray}
\label{equu}
\uu &=&
\left(\overbrace{u_0,u_0,\ldots,u_0}^{s_0},\overbrace{u_1,u_1,\ldots,u_1}^{s_1},\ldots,
\overbrace{u_{t-1},u_{t-1},\ldots,u_{t-1}}^{s_{t-1}},\overbrace{u_t,u_t,\ldots,u_t}^{s_{t}}\right).
\end{eqnarray}

Consider a set $\{\C_i\}$ of $t$ nested $[n,n-u_i,u_i+1]$, $0\leq i\leq t-1$,
RS codes with elements in a finite field
$GF(q)$, $q>\max\{m,n\}$, such that
a parity-check matrix for $\C_i$ is given by

\begin{eqnarray}
\label{Hi}
\hspace{-5mm}
H_{u_i} &\hspace{-2mm}=\hspace{-2mm}&
\left(
\begin{array}{ccccc}
1&1&1&\ldots &1\\
1&\al&\al^2&\ldots &\al^{n-1}\\
1&\al^2&\al^4&\ldots &\al^{2(n-1)}\\
\vdots &\vdots &\vdots &\ddots &\vdots \\
1&\al^{u_i-1}&\al^{2(u_i-1)}&\ldots &\al^{(u_i-1)(n-1)}\\
\end{array}
\right) 
\end{eqnarray}
where $\al$ is an element of order $\cO(\al)\geq \max\{m,n\}$ in
$GF(q)$. Assume also that $\C_t\eq\{0\}$.

Let $\C(n,\uu)$ be the code
consisting of $m\times n$ arrays over
$GF(q)$ such that, for each array in the code with rows
$\uc_0,\uc_1,\ldots,\uc_{m-1}$, $\uc_j\in\C_0$ for $0\leq
j\leq m-1$ and, if 

\begin{eqnarray}
\label{hsi}
\hs_i &=&
\sum_{j=i}^{t}s_j\quad {\rm for}\quad 0\leq i\leq t,
\end{eqnarray}
then
\begin{eqnarray}
\label{eqGP1}
\bigoplus_{j=0}^{m-1}\al^{rj}\uc_j&\in& \C_{t-i}\;\;{\rm for}\;\; 0\leq
i\leq t-1\;\;{\rm and}\;\; 0\leq r\leq \hs_{t-i}-1.
\end{eqnarray}

Then we say that $\C(n,\uu)$ is a $t$-level Extended Integrated Interleaved (EII) code.
\qed
}
\end{definition}

In Definition~\ref{defGPMDS}, notice that, 
if $u_0\eq 0$, then the code $\C_0$ is an
$[n,n,1]$ code, that is, the whole space, with no erasure-correcting
capabilities. Let us mention the recent work in~\cite{z2} on II
codes, in which the definition is modified to improve the locality
when only one row has erasures, but the number of erasures in such
row exceeds $u_0$.

Before giving the properties of $t$-level EII codes, we present some examples.

\begin{example}
\label{ex0}
{\em
Assume that $s_t=0$ in Definition~\ref{defGPMDS}, then, 
in~(\ref{eqGP1}), $i\geq 1$ and
$\C(n,\uu)$ is a $t$-level II code~\cite{bh,tk,w}.
So, $t$-level II codes can be viewed as a special case of $t$-level EII
codes.
}
\end{example}

In~\cite{tk,w}, when $t>2$, II codes are called Generalized
Integrated Interleaved (GII) codes, while II codes
refer to the case $t\eq 2$. The reason for this denomination is
historical, since the first paper on II codes~\cite{hapkt}
describes the case $t\eq 2$ only.

\begin{example}
\label{ex4}
{\em
Assume that $t=1$, then~(\ref{equu}) gives $\uu=
\left(\overbrace{u_0,u_0,\ldots,u_0}^{s_0},\overbrace{n,n,\ldots,n}^{s_1}\right)$.
If $s_1>0$,
$\C(n,\uu)$ is a regular product code~\cite{ms} such that each row
is in an $[n,n-u_0]$ code and each column in an $[m,m-s_1]$ code.
Thus, product codes can be viewed as a special case of $t$-level EII codes.

}
\end{example}

\begin{example}
\label{ex5}
{\em
Assume that $t=2$. Then, $\C_1\subset\C_0$, $$\uu=
\left(\overbrace{u_0,u_0,\ldots,u_0}^{s_0},\overbrace{u_1,u_1,\ldots,u_1}^{s_1},\overbrace{n,n,\ldots,n}^{s_2}\right),$$
$s_0+s_1+s_2=m$,
and consider the 2-level EII code $\C(n,\uu)$. Let $\uc=(\uc_0,\uc_1,\ldots,\uc_{m-1})$ be an
$m\times n$ array in $\C(n,\uu)$ . Then, $\uc_j\in\C_0$ for each
$0\leq j\leq m-1$, 
and, since $\C_2\eq\{0\}$, (\ref{eqGP1}) gives

\begin{eqnarray}
\label{eqGP2e2}
\bigoplus_{j=0}^{m-1}\al^{rj}\uc_j&=& 0\;\;{\rm for}\;\; 0\leq
r\leq s_2-1\\
\label{eqGP1e2}
\bigoplus_{j=0}^{m-1}\al^{rj}\uc_j&\in& \C_{1}\;\;{\rm for}\;\;0\leq r\leq s_1+s_2-1
\end{eqnarray}

The 2-level II codes presented in~\cite{hapkt} correspond to
$s_2= 0$ in this example, i.e., only equations (\ref{eqGP1e2})
are taken into account.

As a special case of 2-level EII codes, take
$$\uu=\left(\overbrace{1,1,\ldots,1}^{m-2},2,n\right)$$ (hence, $s_0\eq m-2$,
$s_1\eq s_2\eq 1$).
The rows $\uc_0,\uc_1,\ldots,\uc_{m-1}$ of
$\C(n,\uu)$
constitute a 2-level II code. Each column is in an $[m,m-1,2]$ code and
each row is in an $[n,n-1,2]$ code (single parity). The $\C_0$ code
is the $[n,n-1,2]$ code, and the $\C_1$ code is the $[n,n-2,3]$ code
given, according to~(\ref{Hi}), by the parity-check matrix

\begin{eqnarray*}
H_1 &=&
\left(
\begin{array}{ccccc}
1&1&1&\ldots &1\\
1&\al&\al^2&\ldots &\al^{n-1}\\
\end{array}
\right).
\end{eqnarray*}
Moreover, (\ref{eqGP2e2}) and~(\ref{eqGP1e2}) give

\begin{eqnarray*}
\bigoplus_{i=0}^{m-1}\uc_i&=&0\\
\bigoplus_{i=0}^{m-1}\al^i\uc_i&\in &\C_1.
\end{eqnarray*}

It is not hard to prove directly that this code can correct any 5
erasures, but this will be a consequence of Theorem~\ref{cor11}
to be presented below. It consists of a
product code (which has minimum distance 4) plus one extra parity.
This extra parity brings the minimum distance up from 4 to 6.
For instance, if $m=4$ and $n=5,$
erasure patterns like the following (vertices of a rectangle)
$$
\begin{array}{|c|c|c|c|c|}
\hline
\phantom{X}&\phantom{X}&\phantom{X}&\phantom{X}&\phantom{X}\\
\hline
\phantom{X}&E&\phantom{X}&\phantom{X}&E\\
\hline
\phantom{X}&\phantom{X}&\phantom{X}&\phantom{X}&\phantom{X}\\
\hline
\phantom{X}&E&\phantom{X}&\phantom{X}&E\\
\hline
\end{array}
$$
are uncorrectable by the product code but can be corrected by
$\C(5,(1,1,2,5))$. An extra erasure in addition to the four
depicted above can be corrected by either the horizontal or the
vertical code.
}
\end{example}

\begin{example}
\label{ex6}
{\em
Assume that $t=3$. Then, $\C_2\subset\C_1\subset\C_0$, $$\uu\hspace{-.5mm}=\hspace{-.5mm}
\left(\hspace{-.5mm}\overbrace{u_0,\hspace{-.3mm}u_0,\hspace{-.3mm}\ldots,\hspace{-.3mm}u_0}^{s_0},
\overbrace{u_1,\hspace{-.3mm}u_1,\hspace{-.3mm}\ldots,\hspace{-.3mm}u_1}^{s_1},
\overbrace{u_2,\hspace{-.3mm}u_2,\hspace{-.3mm}\ldots,u_2}^{s_2},\hspace{-.3mm}
\overbrace{n,\hspace{-.3mm}n,\hspace{-.3mm}\ldots,\hspace{-.3mm}n}^{s_3}\hspace{-1mm}\right),$$
$s_0+s_1+s_2+s_3=m$,
and consider the 3-level EII code $\C(n,\uu)$. Let
$\uc=(\uc_0,\uc_1,\ldots,\uc_{m-1})$ be an
$m\times n$ array in $\C(n,\uu)$. Then, $\uc_j\in\C_0$ for each
$0\leq j\leq m-1$, and 
(\ref{eqGP1}) gives

\begin{eqnarray}
\label{eqGP2e3}
\bigoplus_{j=0}^{m-1}\al^{rj}\uc_j&=& 0\;\;{\rm for}\;\; 0\leq
r\leq s_3-1\\
\label{eqGP1e3}
\bigoplus_{j=0}^{m-1}\al^{rj}\uc_j&\in& \C_{2}\;\;{\rm for}\;\;0\leq
r\leq s_2+s_3-1\\
\label{eqGP1e3b}
\bigoplus_{j=0}^{m-1}\al^{rj}\uc_j&\in& \C_{1}\;\;{\rm for}\;\;0\leq
r\leq s_1+s_2+s_3-1
\end{eqnarray}

}
\end{example}

In Definition~\ref{defGPMDS} of EII codes, we have assumed that the
nested codes are RS codes.
In~\cite{w}, the construction of II codes was adapted to include binary
BCH codes. 
In order to replace the component codes by binary BCH codes, the
$\uc_i$s cannot be multiplied by powers of $\al$ in~(\ref{eqGP1}),
since doing so would take us out 
of the binary field. This problem is overcome in~\cite{w} by replacing the powers of
$\al$ by powers of $x$ modulo the primitive polynomial that defines the
finite field $GF(q)$. 
In future work, we will show how to adapt EII codes to any arbitrary
set of nested codes.

Although the nested codes in
Definition~\ref{defGPMDS} are RS codes, they can be other types of
MDS codes as well, like Extended RS codes~\cite{ms} or Blaum-Roth
(BR)~\cite{br} codes. For simplicity, we concentrate on RS codes. 

\subsection{Erasure Decoding of EII Codes}
\label{EDecod}

We are now ready to state the main result regarding EII codes.

\begin{theorem}
\label{theo2}
{\em
Consider an $m\times n$ array corresponding to a $\C(n,\uu)$
$t$-level EII code as given by Definition~\ref{defGPMDS}. Then, the code
can correct up to $u_0$ erasures in any row and up
to $u_i$ erasures in any $s_i$ rows, where $1\leq i\leq t$.
}
\end{theorem}

\noindent\pf
We may assume that the rows with erasures contain more than $u_0$
erasures, since each row is in $\C_0$, which is an $[n,n-u_0,u_0+1]$
code, hence, 
rows with up to $u_0$ erasures can be corrected.

Assume that there are
$\ell$ rows with more than $u_0$ erasures such that there are up
to $u_i$ erasures in any up to $s_i$ rows, $1\leq i\leq t$.
We do induction on $\ell$.

If $\ell=0$, there is nothing to prove,
so, assume that there are $\ell\geq 1$ rows with more than $u_0$
erasures each
such that there are up
to $u_i$ erasures in any up to $s_i$ rows, $1\leq i\leq t$.
In particular, $\ell\leq s_1+s_2+\ldots +s_{t}\eq\hs_1$.
By induction, up to $\ell -1$ rows with this property are correctable.

Let $i_0,i_1,\ldots,i_{m-1}$
be an ordering of the rows according to a
non-increasing number of erasures such that:

\begin{enumerate}

\item Row $i_{j}$ for $0\leq
j\leq \ell-1$ has $v_j$ erasures, where \\$n\geq v_0\geq v_1\geq
\ldots\geq v_{\ell-1}>u_0$.

\item Rows $i_{\ell},i_{\ell+1},\ldots,i_{m-1}$ have no erasures.

\end{enumerate}

It suffices to prove that the $v_{\ell-1}$ erasures in row $i_{\ell-1}$ can be
corrected. Then we will be left with $\ell -1$ rows with more than $s_0$
erasures each such that there are up
to $u_i$ erasures in any up to $s_i$ rows, $1\leq i\leq t$,
and the result
follows by induction.

In effect, define $w$, $1\leq w\leq t-1$, such that

\begin{eqnarray}
\label{w}
\hs_{w+1}\eq \sum_{i=1}^{t-w}s_{w+i}
<\ell\leq
\sum_{i=0}^{t-w}s_{w+i}\eq \hs_w 
\end{eqnarray}
and consider the
code $\C_w$ from the nested set of codes $\C_i$ in
Definition~\ref{defGPMDS}, which can correct up to $u_w$ erasures.
Since there are up to $u_w$ erasures in any up to $s_w$ rows, given
that the $v_j$s are non-increasing and $0<\ell-\hs_{w+1}\leq
s_w$, then $v_j\leq u_w$ for $\hs_{w+1}\leq
j\leq \ell-1$. In particular, $v_{\ell-1}\leq u_w$.

Rearranging the order of the elements of the sums
in~(\ref{eqGP1}), and since
$\C_{t}\subset\C_{t-1}\subset\cdots\subset\C_w$,
from~(\ref{eqGP1}), in particular, we have

\begin{eqnarray}
\label{eqGP11}
\bigoplus_{j=0}^{m-1}\,\al^{ri_j}\,\uc_{i_j}
&\in& \C_{w}\;\;{\rm for}\;\; 0\leq r\leq \ell -1.
\end{eqnarray}
Since the $\ell\times m$
matrix corresponding to the
coefficients of the $\uc_{i_j}$s
in~(\ref{eqGP11}) is a Vandermonde type of matrix and $\cO(\al)\geq
\max\{m,n\}$, 
this matrix can be triangulated, giving

\begin{eqnarray}
\label{eqGP11t}
\hspace{-5mm}
\uc_{i_r}\xor
\left(\bigoplus_{j=r+1}^{m-1}\,\ga_{r,j}\,\uc_{i_j}\right)
&\in& \C_{w}\;\;{\rm for}\;\; 0\leq r\leq \ell -1,
\end{eqnarray}
where the coefficients $\ga_{r,j}$ are a result of the triangulation.
In particular, taking $r=\ell -1$ in~(\ref{eqGP11t}), we
obtain

\begin{eqnarray}
\label{eqGP11tl}
\uc_{i_{\ell -1}}
\xor
\left(\bigoplus_{j=\ell}^{m-1}\,\ga_{\ell -1,j}\,\uc_{i_j}\right)
&\in& \C_{w}.
\end{eqnarray}
Since $\uc_{i_{\ell -1}}$ has $v_{\ell-1}$ erasures and
$\uc_{i_j}$ has no erasures for \\$\ell\leq j\leq m-1$, then
$\uc_{i_{\ell -1}}
\xor
\left(\bigoplus_{j=\ell}^{m-1}\ga_{\ell
-1,j}\,\uc_{i_j}\right)$ has $v_{\ell-1}$ erasures. Since the vector is
in $\C_w$ and $v_{\ell-1}\leq u_w$, the erasures can be corrected.
Once $\uc_{i_{\ell -1}}
\xor
\left(\bigoplus_{j=\ell}^{m-1}\ga_{\ell
-1,j}\,\uc_{i_j}\right)$ is corrected, $\uc_{i_{\ell -1}}$ is
obtained as

\begin{eqnarray*}
\uc_{i_{\ell -1}}&=&
\left(\uc_{i_{\ell -1}}\xor
\left(\bigoplus_{j=\ell}^{m-1}\ga_{\ell
-1,j}\,\uc_{i_j}\right)\right)\xor\left(\bigoplus_{j=\ell}^{m-1}\ga_{\ell
-1,j}\,\uc_{i_j}\right)
\end{eqnarray*}
and the result follows by induction on $\ell$.
\qed

Theorem~\ref{theo2} generalizes Theorem~1 in~\cite{bh}.
The proof of Theorem~\ref{theo2} is constructive in the sense that it
provides a decoding algorithm. The following example illustrates
Theorem~\ref{theo2} and the decoding algorithm.

\begin{example}
\label{ex7}
{\em
Consider the 3-level EII code $\C(7,(1,1,3,4,7,7))$ according to
Definition~\ref{defGPMDS} and Example~\ref{ex6}.
We have four codes $\C_3\subset\C_2\subset\C_1\subset\C_0$, where
$\C_0$ is a $[7,6,2]$ code, $\C_1$ is a $[7,4,4]$ code, $\C_2$ is
a $[7,3,5]$ code and $\C_3\eq\{0\}$. 
We may assume that the entries of these codes are
in $GF(8)$ and that $\al$ is a primitive element in $GF(8)$.

Consider the following $6\times 7$ array with erasures denoted by $E$:

$$
\begin{array}{rl}
\begin{array}{c}
\uc_0\\\uc_1\\\uc_2\\\uc_3\\\uc_4\\\uc_5\\
\end{array}
&
\begin{array}{|c|c|c|c|c|c|c|}
\hline
\phantom{X}&\phantom{X}&E&\phantom{X}&\phantom{X}&\phantom{X}&\phantom{X}\\
\hline
E&E&E&E&E&E&E\\
\hline
\phantom{X}&E&E&\phantom{X}&E&\phantom{X}&E\\
\hline
E&\phantom{X}&\phantom{X}&E&\phantom{X}&E&\phantom{X}\\
\hline
E&E&E&E&E&E&E\\
\hline
\phantom{X}&\phantom{X}&\phantom{X}&\phantom{X}&\phantom{X}&E&\phantom{X}\\
\hline
\end{array}
\end{array}
$$

The first step is correcting the single erasures in $\uc_0$ and in
$\uc_5$. An ordering of the remaining rows in non-increasing number
of erasures is $\{i_0,i_1,i_2,i_3\}=\{1,4,2,3\}$ ($\ell\eq 4$). In particular,
$\uc_3$ has three erasures.
According to~(\ref{eqGP2e3}), (\ref{eqGP1e3}) and~(\ref{eqGP1e3b}),

$$
\begin{array}{ccccccccccccl}
\uc_0&\hspace{-3mm}\xor\hspace{-3mm}
&\uc_1&\hspace{-3mm}\xor\hspace{-3mm} &\uc_2&
\hspace{-3mm}\xor\hspace{-3mm} &\uc_3&\hspace{-3mm}\xor\hspace{-3mm}  &\uc_4&\hspace{-3mm}\xor\hspace{-3mm} &\uc_5&= &0\\
\uc_0&\hspace{-3mm}\xor\hspace{-3mm} &\al
\uc_1&\hspace{-3mm}\xor\hspace{-3mm}
&\al^2\uc_2&\hspace{-3mm}\xor\hspace{-3mm}  &\al^3\uc_3&\hspace{-3mm}\xor\hspace{-3mm} &\al^4\uc_4&\hspace{-3mm}\xor\hspace{-3mm} &\al^5\uc_5&=&0\\
\uc_0&\hspace{-3mm}\xor\hspace{-3mm}
&\al^2\uc_1&\hspace{-3mm}\xor\hspace{-3mm}
&\al^4\uc_2&\hspace{-3mm}\xor\hspace{-3mm}
&\al^6\uc_3&\hspace{-3mm}\xor\hspace{-3mm}  &\al^8\uc_4&\hspace{-3mm}\xor\hspace{-3mm}  &\al^{10}\uc_5&\in &\C_2\\
\uc_0&\hspace{-3mm}\xor\hspace{-3mm}
&\al^3\uc_1&\hspace{-3mm}\xor\hspace{-3mm}
&\al^6\uc_2&\hspace{-3mm}\xor\hspace{-3mm}
&\al^9\uc_3&\hspace{-3mm}\xor\hspace{-3mm}  &\al^{12}\uc_4& \hspace{-3mm}\xor\hspace{-3mm} &\al^{15}\uc_5&\in &\C_1.\\
\end{array}
$$

Notice that $\C_1$ can correct three erasures, i.e., $w=1$ in~(\ref{w}).
Rearranging the $\uc_i$s above in non-increasing number
of erasures, we obtain

$$
\begin{array}{ccccccccccccl}
\uc_1&\hspace{-3mm}\xor\hspace{-3mm}
&\uc_4&\hspace{-3mm}\xor\hspace{-3mm}
&\uc_2&\hspace{-3mm}\xor\hspace{-3mm} &\uc_3&  
\hspace{-3mm}\xor\hspace{-3mm} &\uc_0&\hspace{-3mm}\xor\hspace{-3mm} &\uc_5&= &0\\
\al \uc_1&\hspace{-3mm}\xor\hspace{-3mm}
&\al^4\uc_4&\hspace{-3mm}\xor\hspace{-3mm}
&\al^2\uc_2&\hspace{-3mm}\xor\hspace{-3mm} &\al^3\uc_3&
\hspace{-3mm}\xor\hspace{-3mm} & \uc_0&\hspace{-3mm}\xor\hspace{-3mm} &\al^5\uc_5&=&0\\
\al^2\uc_1&\hspace{-3mm}\xor\hspace{-3mm}
&\al^8\uc_4&\hspace{-3mm}\xor\hspace{-3mm} &
\al^4\uc_2&\hspace{-3mm}\xor\hspace{-3mm} &
\al^6\uc_3&\hspace{-3mm}\xor\hspace{-3mm} & \uc_0&\hspace{-3mm}\xor\hspace{-3mm} &\al^{10}\uc_5&\in &\C_2\\
\al^3\uc_1&\hspace{-3mm}\xor\hspace{-3mm}
&\al^{12}\uc_4&\hspace{-3mm}\xor\hspace{-3mm}
&\al^6\uc_2&\hspace{-3mm}\xor\hspace{-3mm}
&\al^9\uc_3&\hspace{-3mm}\xor\hspace{-3mm} &\uc_0& \hspace{-3mm}\xor\hspace{-3mm} &\al^{15}\uc_5&\in &\C_1,\\
\end{array}
$$
which corresponds to~(\ref{eqGP11}) in the
proof of Theorem~\ref{theo2}.
Triangulating this linear system in $GF(8)$, where $1\xor\al\xor\al^3=0$,
and since $\C_2\subset\C_1$, we obtain the
following triangulated system:

$$
\begin{array}{ccccccccccccl}
\uc_1&\hspace{-3mm}\xor\hspace{-3mm}
&\uc_4&\hspace{-3mm}\xor\hspace{-3mm}
&\uc_2&\hspace{-3mm}\xor\hspace{-3mm} &  
\uc_3&\hspace{-3mm}\xor\hspace{-3mm} &\uc_0&\hspace{-3mm}\xor\hspace{-3mm} &\uc_5&= &0\\
&&\uc_4&\hspace{-3mm}\xor\hspace{-3mm}
&\al^2\uc_2&\hspace{-3mm}\xor\hspace{-3mm}  
&\al^5\uc_3&\hspace{-3mm}\xor\hspace{-3mm} &\al
\uc_0&\hspace{-3mm}\xor\hspace{-3mm}  &\al^4\uc_5&=&0\\
&&&&\uc_2&\hspace{-3mm}\xor\hspace{-3mm} &\al
\uc_3&\hspace{-3mm}\xor\hspace{-3mm}  &\al^3\uc_0&\hspace{-3mm}\xor\hspace{-3mm} &\al \uc_5&\in &\C_2\\
&&&&&&\uc_3&\hspace{-3mm}\xor\hspace{-3mm}
&\al^3\uc_0&\hspace{-3mm}\xor\hspace{-3mm}  &\al^{5}\uc_5&\in &\C_1.\\
\end{array}
$$

Since $\uc_3$ has 3 erasures and $\uc_0$ and $\uc_5$ have no
erasures, $\uc_3\xor\al^3\uc_0\xor\al^{5}\uc_5$ has 3
erasures, which can be corrected in $\C_1$.
Then, $$\uc_3=(\uc_3\xor\al^3\uc_0\xor\al^{5}\uc_5)\xor
(\al^3\uc_0\xor\al^{5}\uc_5).$$

Similarly, $\uc_2\xor\al \uc_3\xor\al^3\uc_0\xor\al \uc_5$ has 4
erasures, which can be corrected in $\C_2$,
and $$\uc_2=(\uc_2\xor\al \uc_3\xor\al^3\uc_0\xor\al \uc_5)\xor
(\al \uc_3\xor\al^3\uc_0\xor\al \uc_5).$$

Finally, since the first two rows of the triangulated system are
equal to zero, we obtain

\begin{eqnarray*}
\uc_4&=&\al^2\uc_2\xor\al^5\uc_3\xor\al \uc_0\xor\al^4\uc_5\\
\uc_1&=&\uc_4\xor\uc_2\xor\uc_3\xor\uc_0\xor\uc_5,
\end{eqnarray*}
completing the decoding.
}
\end{example}

From the proof of Theorem~\ref{theo2}, even if the decoding algorithm
cannot correct all the erasures, it is often possible to correct a
few rows. Specifically, consider a $\C(n,\uu)$
$t$-level EII code as given by Definition~\ref{defGPMDS}, and assume
that, given a received array in $\C(n,\uu)$, row $i$ of the array has $x_i$
erasures. Let $x_{i_0}\leq x_{i_1}\leq \cdots\leq x_{i_{m-1}}$ and
consider $\uu$ as given by~(\ref{equu}). Let $\uu\eq
(v_0,v_1,\ldots,v_{m-1})$, where, by~(\ref{equu}), $v_0\leq
v_1\leq\cdots\leq v_{m-1}$. Define $y$, $0\leq y\leq m-1$, such
that $x_{i_j}\leq v_j$ for $0\leq j\leq y$, and
$x_{y+1}>v_{y+1}$. If $y\eq m-1$, 
then, by Theorem~\ref{theo2}, all the erasures in the array are
correctable. However, if $y< m-1$, then rows $i_0,i_1,\ldots,i_y$
are still correctable by the triangulation algorithm, but rows $i_j$
for $y+1\leq j\leq m-1$ are not. We illustrate this fact in the
following example.

\begin{example}
\label{ex7bis}
{\em
Consider the 4-level EII code $\C(7,(1,2,3,5))$ according to
Definition~\ref{defGPMDS}.
We have four codes $\C_3\subset\C_2\subset\C_1\subset\C_0$, where
$\C_0$ is a $[7,6,2]$ code, $\C_1$ is a $[7,5,3]$ code, $\C_2$ is
a $[7,4,4]$ code and $\C_3$ is a $[7,2,6]$ code. 
As in Example~\ref{ex7}, we assume that the entries of these codes are
in $GF(8)$ and that $\al$ is a primitive element in $GF(8)$.

Consider the following $4\times 7$ array with erasures denoted by $E$:

$$
\begin{array}{rl}
\begin{array}{c}
\uc_0\\\uc_1\\\uc_2\\\uc_3\\
\end{array}
&
\begin{array}{|c|c|c|c|c|c|c|}
\hline
E&\phantom{X}&\phantom{X}&E&\phantom{X}&E&E\\
\hline
&E&&E&&&\\
\hline
&&E&&&&\\
\hline
E&E&&&&E&E\\
\hline
\end{array}
\end{array}
$$

Then, according to the notation above, $x_0\eq 4$, $x_1\eq 2$,
$x_2\eq 1$ and $x_3\eq 4$. Writing this in non-decreasing order, we
have $x_2\leq x_1\leq x_0\leq x_3$. Since $x_2\leq v_0$, $x_1\leq
v_1$ and $x_0>v_2$, $y\eq 1$. Since $\uc_2$ has a single erasure,
we may assume that this erasure is corrected in $\C_0$, so now
$\uc_2$ is erasure free. According to Definition~\ref{defGPMDS},
we have the following system:

$$
\begin{array}{ccccccccl}
\uc_0&\xor &\uc_3&\xor &\uc_1&\xor &\uc_2&\in &\C_3\\
\uc_0&\xor &\al^3\uc_3&\xor &\al\uc_1&\xor &\al^2\uc_2&\in &\C_2\\
\uc_0&\xor &\al^6\uc_3&\xor &\al^2\uc_1&\xor &\al^4\uc_2&\in &\C_1\\
\end{array}
$$

Triangulating this system, we obtain

$$
\begin{array}{ccccccccl}
\uc_0&\xor &\uc_3&\xor &\uc_1&\xor &\uc_2&\in &\C_3\\
& &\uc_3&\xor &\al^2\uc_1&\xor &\al^5\uc_2&\in &\C_2\\
& && &\uc_1&\xor &\al\uc_2&\in &\C_1\\
\end{array}
$$

Since $\uc_1\xor\al\uc_2$ has two erasures, they can be corrected
in $\C_1$. Then, $\uc_1$ is obtained as $\uc_1\eq
(\uc_1\xor\al\uc_2)\xor \al\uc_2$. However, with this procedure,
$\uc_3\xor\al^2\uc_1\xor\al^5\uc_2$ cannot be obtained, since 4
erasures are uncorrectable in $\C_2$, so after correcting $\uc_1$ and
$\uc_2$, we are left with the uncorrectable array

$$
\begin{array}{rl}
\begin{array}{c}
\uc_0\\\uc_1\\\uc_2\\\uc_3\\
\end{array}
&
\begin{array}{|c|c|c|c|c|c|c|}
\hline
E&\phantom{X}&\phantom{X}&E&\phantom{X}&E&E\\
\hline
&\phantom{X}&&\phantom{X}&&&\\
\hline
&&\phantom{X}&&&&\\
\hline
E&E&&&&E&E\\
\hline
\end{array}
\end{array}
$$

}
\end{example}

Although the erasure pattern in Example~\ref{ex7bis} is only
partially correctable, we will see after Theorem~\ref{theo3} that it
can be fully correctable when expanding the correction to columns.

\subsection{Dimension, Encoding and Minimum Distance of EII Codes}
\label{dimdist}

Before discussing the dimension, the encoding and the minimum
distance of $t$-level EII codes, let us state and prove the following lemma.

\begin{lemma}
\label{lemma1}
{\em
Consider the $t$-level EII code $\C(n,\uu)$ as given
by Definition~\ref{defGPMDS}. 
Then, for each $j$ such that $0\leq j\leq t-1$, given $u_j+1$ fixed
column indices in $\hs_{j+1}+1$
different rows, there is an array in $\C(n,\uu)$ that is
non-zero in such $\left(\hs_{j+1}+1\right)\left(u_j+1\right)$
locations and 0 elsewhere.
}
\end{lemma}

\noindent\pf
Since $\C_j$ is an $[n,n-u_j,u_j\nobreak +1]$ MDS code for each $j$
such that $0\leq j\leq t-1$, given $u_j+1$ fixed indices in a 
vector of length $n$, there is a codeword
$\uw$ in $\C_j$ whose non-zero entries are in such $u_j+1$ fixed
locations. Assume that the $\hs_{j+1}+1$ rows selected are
$i_0,i_1,\ldots,i_{\hs_{j+1}}$, where $$0\leq i_0<i_1 <\ldots
<i_{\hs_{j+1}}\leq m-1.$$
Let $\uv=(v_0,v_1,\ldots,v_{\hs_{j+1}})$ be a vector of weight $\hs_{j+1}+1$
such that

\begin{eqnarray}
\label{ref}
\bigoplus_{s=0}^{\hs_{j+1}}\al^{ri_s}v_s&=&0\;\;{\rm for}\;\;0\leq
r\leq \hs_{j+1}-1.
\end{eqnarray}

Such a vector exists since the coefficients in~(\ref{ref}) are in an
$\hs_{j+1}\times (\hs_{j+1}+1)$ Vandermonde matrix (which corresponds
to the parity-check matrix of an $[\hs_{j+1}+1,1,\hs_{j+1}+1]$ RS code).
Consider the $m\times n$ array of weight
$\left(\hs_{j+1}+1\right)\left(u_j+1\right)$ such that row $i_s$ equals
$v_s\,\uw$ for $0\leq s\leq \hs_{j+1}$, and the remaining rows are zero.
We will show that this array is in
$\C(n,\uu)$ . Since each row of the array is in $\C_j$ by design, in
particular, it is in $\C_0$.
According to~(\ref{eqGP1}), 
we have to show that

\begin{eqnarray}
\label{eqGP11b}
\bigoplus_{s=0}^{\hs_{j+1}}\al^{ri_s}\left(v_s\,\uw\right)&\in&
\C_{t-i}\;\;{\rm for}\;\; 0\leq i\leq t-1\;\;
{\rm and}\;\;0\leq r\leq \hs_{t-i}-1.
\end{eqnarray}

If $0\leq j\leq t-i-1$, then, $\hs_{j+1}\geq \hs_{t-i}$,
and, for $0\leq r\leq \hs_{t-i}-1$, by~(\ref{ref}),
\begin{eqnarray*}
\bigoplus_{s=0}^{\hs_{j+1}}\al^{ri_s}\left(v_s\,\uw\right)\;\;=\;\;
\left(\bigoplus_{s=0}^{\hs_{j+1}}\al^{ri_s}v_s\right)\,\uw&=
&0,\end{eqnarray*}
so, in particular, 
(\ref{eqGP11b}) follows.

If $t-i\leq j\leq t-1$, then $\C_j\subseteq
\C_{t-i}$ and $\uw\in\C_{t-i}$,
so~(\ref{eqGP11b}) also follows in this case.
\qed

\begin{example}
\label{ex9}
{\em
Consider the 3-level EII code $\C(7,\hspace{-.3mm}(1,\hspace{-.2mm}1,
\hspace{-.2mm}3,\hspace{-.2mm}4,\hspace{-.2mm}7,\hspace{-.2mm}7)\hspace{-.3mm})$ of
Example~\ref{ex7}. According to Lemma~\ref{lemma1}, the locations
denoted by $E$ in the following arrays correspond to the non-zero
entries of arrays in $\C(7,(1,1,3,4,7,7))$
for $j =0$, 1 and 2 respectively:
$$
\begin{array}{ll}
\begin{array}{|c|c|c|c|c|c|c|}
\hline
\hspace{-.2mm}\phantom{E}\hspace{-.2mm}&\hspace{-.2mm}E\hspace{-.2mm}&\hspace{-.2mm}\phantom{E}\hspace{-.2mm}
&\hspace{-.2mm}E\hspace{-.2mm}&\hspace{-.2mm}\phantom{E}\hspace{-.2mm}&
\hspace{-.2mm}\phantom{E}\hspace{-.2mm}&\hspace{-.2mm}\phantom{E}\hspace{-.2mm}\\
\hline
\hspace{-.2mm}\phantom{E}\hspace{-.2mm}&\hspace{-.2mm}E\hspace{-.2mm}&\hspace{-.2mm}\phantom{E}\hspace{-.2mm}
&\hspace{-.2mm}E\hspace{-.2mm}&\hspace{-.2mm}\phantom{E}\hspace{-.2mm}&\hspace{-.2mm}\phantom{E}\hspace{-.2mm}&
\hspace{-.2mm}\phantom{E}\hspace{-.2mm}\\
\hline
\hspace{-.2mm}\phantom{E}\hspace{-.2mm}&\hspace{-.2mm}\phantom{E}\hspace{-.2mm}&
\hspace{-.2mm}\phantom{E}\hspace{-.2mm}&\hspace{-.2mm}\phantom{E}\hspace{-.2mm}&\hspace{-.2mm}\phantom{E}\hspace{-.2mm}
&\hspace{-.2mm}\phantom{E}\hspace{-.2mm}&\hspace{-.2mm}\phantom{E}\hspace{-.2mm}\\
\hline
\hspace{-.2mm}\phantom{E}\hspace{-.2mm}&\hspace{-.2mm}E\hspace{-.2mm}&\hspace{-.2mm}\phantom{E}\hspace{-.2mm}
&\hspace{-.2mm}E\hspace{-.2mm}&\hspace{-.2mm}\phantom{E}\hspace{-.2mm}&
\hspace{-.2mm}\phantom{E}\hspace{-.2mm}&\hspace{-.2mm}\phantom{E}\hspace{-.2mm}\\
\hline
\hspace{-.2mm}\phantom{E}\hspace{-.2mm}&\hspace{-.2mm}E\hspace{-.2mm}&
\hspace{-.2mm}\phantom{E}\hspace{-.2mm}&\hspace{-.2mm}E\hspace{-.2mm}&
\hspace{-.2mm}\phantom{E}\hspace{-.2mm}&\hspace{-.2mm}\phantom{E}\hspace{-.2mm}&\hspace{-.2mm}\phantom{E}\hspace{-.2mm}\\
\hline
\hspace{-.2mm}\phantom{E}\hspace{-.2mm}&\hspace{-.2mm}E\hspace{-.2mm}&
\hspace{-.2mm}\phantom{E}\hspace{-.2mm}&\hspace{-.2mm}E\hspace{-.2mm}&\hspace{-.2mm}\phantom{E}\hspace{-.2mm}&
\hspace{-.2mm}\phantom{E}\hspace{-.2mm}&\hspace{-.2mm}\phantom{E}\hspace{-.2mm}\\
\hline
\end{array}
&
\begin{array}{|c|c|c|c|c|c|c|}
\hline
\hspace{-.2mm}\phantom{E}\hspace{-.2mm}&\hspace{-.2mm}E\hspace{-.2mm}&
\hspace{-.2mm}E\hspace{-.2mm}&\hspace{-.2mm}\phantom{E}\hspace{-.2mm}&
\hspace{-.2mm}E\hspace{-.2mm}&\hspace{-.2mm}\phantom{E}\hspace{-.2mm}&\hspace{-.2mm}E\hspace{-.2mm}\\
\hline
\hspace{-.2mm}\phantom{E}\hspace{-.2mm}&\hspace{-.2mm}\phantom{E}\hspace{-.2mm}&
\hspace{-.2mm}\phantom{E}\hspace{-.2mm}&\hspace{-.2mm}\phantom{E}\hspace{-.2mm}&
\hspace{-.2mm}\phantom{E}\hspace{-.2mm}&\hspace{-.2mm}\phantom{E}\hspace{-.2mm}&\hspace{-.2mm}\phantom{E}\hspace{-.2mm}\\
\hline
\hspace{-.2mm}\phantom{E}\hspace{-.2mm}&\hspace{-.2mm}E\hspace{-.2mm}&
\hspace{-.2mm}E\hspace{-.2mm}&\hspace{-.2mm}\phantom{E}\hspace{-.2mm}&
\hspace{-.2mm}E\hspace{-.2mm}&\hspace{-.2mm}\phantom{E}\hspace{-.2mm}&\hspace{-.2mm}E\hspace{-.2mm}\\
\hline
\hspace{-.2mm}\phantom{E}\hspace{-.2mm}&\hspace{-.2mm}E\hspace{-.2mm}&
\hspace{-.2mm}E\hspace{-.2mm}&\hspace{-.2mm}\phantom{E}\hspace{-.2mm}&\hspace{-.2mm}E\hspace{-.2mm}&
\hspace{-.2mm}\phantom{E}\hspace{-.2mm}&\hspace{-.2mm}E\hspace{-.2mm}\\
\hline
\hspace{-.2mm}\phantom{E}\hspace{-.2mm}&\hspace{-.2mm}\phantom{E}\hspace{-.2mm}&\hspace{-.2mm}\phantom{E}\hspace{-.2mm}
&\hspace{-.2mm}\phantom{E}\hspace{-.2mm}&\hspace{-.2mm}\phantom{E}\hspace{-.2mm}&
\hspace{-.2mm}\phantom{E}\hspace{-.2mm}&\hspace{-.2mm}\phantom{E}\hspace{-.2mm}\\
\hline
\hspace{-.2mm}\phantom{E}\hspace{-.2mm}&\hspace{-.2mm}E\hspace{-.2mm}&\hspace{-.2mm}E\hspace{-.2mm}&
\hspace{-.2mm}\phantom{E}\hspace{-.2mm}&\hspace{-.2mm}E\hspace{-.2mm}&\hspace{-.2mm}\phantom{E}\hspace{-.2mm}&\hspace{-.2mm}E\hspace{-.2mm}\\
\hline
\end{array}
\end{array}
$$

$$
\begin{array}{|c|c|c|c|c|c|c|}
\hline
\hspace{-.2mm}\phantom{E}\hspace{-.2mm}&\hspace{-.2mm}\phantom{E}\hspace{-.2mm}&\hspace{-.2mm}\phantom{E}\hspace{-.2mm}
&\hspace{-.2mm}\phantom{E}\hspace{-.2mm}&\hspace{-.2mm}\phantom{E}\hspace{-.2mm}&
\hspace{-.2mm}\phantom{E}\hspace{-.2mm}&\hspace{-.2mm}\phantom{E}\hspace{-.2mm}\\
\hline
\hspace{-.2mm}E&\hspace{-.2mm}\phantom{E}\hspace{-.2mm}&\hspace{-.2mm}E\hspace{-.2mm}&\hspace{-.2mm}\phantom{E}\hspace{-.2mm}
&\hspace{-.2mm}E\hspace{-.2mm}&\hspace{-.2mm}E\hspace{-.2mm}&\hspace{-.2mm}E\hspace{-.2mm}\\
\hline
\hspace{-.2mm}\phantom{E}\hspace{-.2mm}&\hspace{-.2mm}\phantom{E}\hspace{-.2mm}&
\hspace{-.2mm}\phantom{E}\hspace{-.2mm}&\hspace{-.2mm}\phantom{E}\hspace{-.2mm}&
\hspace{-.2mm}\phantom{E}\hspace{-.2mm}&\hspace{-.2mm}\phantom{E}\hspace{-.2mm}&\hspace{-.2mm}\phantom{E}\hspace{-.2mm}\\
\hline
\hspace{-.2mm}\phantom{E}\hspace{-.2mm}&\hspace{-.2mm}\phantom{E}\hspace{-.2mm}&
\hspace{-.2mm}\phantom{E}\hspace{-.2mm}&\hspace{-.2mm}\phantom{E}\hspace{-.2mm}&
\hspace{-.2mm}\phantom{E}\hspace{-.2mm}&\hspace{-.2mm}\phantom{E}\hspace{-.2mm}&\hspace{-.2mm}\phantom{E}\hspace{-.2mm}\\
\hline
\hspace{-.2mm}E&\hspace{-.2mm}\phantom{E}\hspace{-.2mm}&\hspace{-.2mm}E\hspace{-.2mm}&\hspace{-.2mm}\phantom{E}\hspace{-.2mm}&
\hspace{-.2mm}E\hspace{-.2mm}&\hspace{-.2mm}E\hspace{-.2mm}&\hspace{-.2mm}E\hspace{-.2mm}\\
\hline
\hspace{-.2mm}E&\hspace{-.2mm}\phantom{E}\hspace{-.2mm}&\hspace{-.2mm}E\hspace{-.2mm}&
\hspace{-.2mm}\phantom{E}\hspace{-.2mm}&\hspace{-.2mm}E\hspace{-.2mm}&
\hspace{-.2mm}E\hspace{-.2mm}&\hspace{-.2mm}E\hspace{-.2mm}\\
\hline
\end{array}
$$

The arrays with erasures in locations $E$ above are uncorrectable
since, provided the zero array was stored, the decoding cannot decide
between the zero array and the arrays with non-zero entries in the
locations $E$.
}
\end{example}
Next we give an auxiliary general lemma.

\begin{lemma}
\label{lemma6}
{\em
Consider an $[n,k]$ linear code, and let $S=\{i_0,i_1,\ldots,i_{s-1}\}$,
where $0\leq i_0<i_1<\cdots <i_{s-1}\leq n-1$.
Assume that, given a codeword with erasures in $S$, the code can
correct such erasures, while, for any $i\not\in S$, erasures in
$S\cup \{i\}$ are not correctable. Then, $n-k=s$.
}
\end{lemma}

\noindent\pf Since the erasures in $S$ are correctable, there are at
least $s$ linearly independent parity equations, so $n-k\,\geq\,s$.

Assume that $n-k>s$. Let $H$ be an $(n-k)\times n$ parity-check matrix of the code
such that the first $s$ rows of $H$ are used to correct the $s$
erasures in $S$, thus, the $s\times s$ submatrix consisting of those first $s$
rows and columns $i_0,i_1,\ldots,i_{s-1}$ is invertible.

Consider next the matrix consisting of the first $s+1$ rows in $H$.
By row operations, we can make the entries $i_0,i_1,\ldots,i_{s-1}$
in the $(s+1)$-th row equal to zero. Since the first $s+1$ rows of
$H$ have rank $s+1$, then there is a non-zero location $i$, $i\not\in
S$, in the $(s+1)$-th row. Thus, columns $S\cup\{i\}$ in the first
$s+1$ rows of $H$ are linearly
independent and hence erasures in $S\cup\{i\}$ are
correctable, a contradiction, so $n-k=s$.
\qed

\begin{theorem}
\label{cor00}
{\em
Consider the $t$-level EII code $\C(n,\uu)$ as given by
Definition~\ref{defGPMDS}. Then, $\C(n,\uu)$ is an $[mn,k]$ code, where

\begin{eqnarray}
\label{eqK}
k&=&
mn-\left(\sum_{i=0}^{t}s_iu_i\right)
\end{eqnarray}

}
\end{theorem}

\noindent\pf
Assume that the zero array is stored, and a received array $W$ has
erasures in the last $u_i$ entries of rows
$m-\hs_{i}$ to $m-\hs_{i+1}-1$ for $0\leq i\leq t-1$,
and in all the entries of rows $m-s_t$
to $m-1$. Thus, $W$ has a total
of $\sum_{i=0}^{t}s_iu_i$ erasures,
and by Theorem~\ref{theo2}, it will be correctly decoded as the zero codeword.

Consider an array $V$ which coincides with $W$, except in one
location in which it has an extra erasure.
If we show that any such $V$ is
uncorrectable, by Lemma~\ref{lemma6}, 
$mn-k\,=\sum_{i=0}^{t}s_iu_i$,
which is equivalent to~(\ref{eqK}).

For each $(u',v')$ 
that is not in the set of erasures
of $W$, define $i'$, $0\leq i'\leq t-1$, such that $m-\hs_{i'}\leq
u'\leq m-\hs_{i'+1}-1$,
and let $Y^{(u',v')}=
\left(y^{(u',v')}_{a,b}\right)_{\substack{0\leq a\leq m-1\\ 0\leq b\leq n-1}}$
be an array in $\C(n,\uu)$ whose non-zero coordinates
are in the intersection
of rows $u',u'+1,\ldots,u'+\hs_{i'+1}$ 
and columns
$v',n-u_{i'},n-u_{i'}+1,\ldots,n-1$. Such a non-zero array exists
due to Lemma~\ref{lemma1}.

Assume that the extra erasure in $V$ is in location $(u,v)$ and
consider $j$, $1\leq j\leq t$, such that
$n-u_{j}\leq v\leq n-u_{j-1}-1$.
Take the arrays $Y^{(u',v)}$, where $u\leq u'\leq m-\hs_{j}-1$.
For each $u'$, $u<u'\leq m-\hs_{j}-1$, choose constants $c_{u'}$ such that

\begin{eqnarray}
\label{yuv}
y^{(u,v)}_{u',v}\,\oplus\,\bigoplus_{z=u+1}^{u'}c_zy^{(z,v)}_{u',v}&=&0.
\end{eqnarray}

Then, if
$Y=\bigoplus_{z=u}^{m-\hs_j-1}Y^{(z,v)},$ 
by~(\ref{yuv}), $Y$ has a non-zero entry in $(u,v)$, while its remaining non-zero
entries are contained in the locations of the erasures of $W$. So,
array 
$V$ is uncorrectable, since it
can be decoded either as the zero array or as $Y$.
\qed

Theorem II.1 in~\cite{tk}, which corresponds to Corollary~2
in~\cite{bh}, is a special case of Theorem~\ref{cor00}.

\begin{example}
\label{ex10}
{\em
We illustrate the proof of Theorem~\ref{cor00} with the 3-level EII
code $\C(7,(1,1,3,4,7,7))$ of Examples~\ref{ex7} and~\ref{ex9}.
By Theorem~\ref{cor00}, this code is a $[42,19]$ code.
Following the proof of Theorem~\ref{cor00}, denote by $E$ the
erased locations in an array $W$:

\begin{eqnarray*}
W&=&
\begin{array}{|c|c|c|c|c|c|c|}
\hline
\phantom{X}&\phantom{X}&\phantom{X}&\phantom{X}&\phantom{X}&\phantom{X}&\hspace{-.2mm}E\hspace{-.2mm}\\
\hline
\phantom{X}&\phantom{X}&\phantom{X}&\phantom{X}&\phantom{X}&\phantom{X}&E\\
\hline
\phantom{X}&\phantom{X}&\phantom{X}&\phantom{X}&E&E&E\\
\hline
\phantom{X}&\phantom{X}&\phantom{X}&E&E&E&E\\
\hline
E&E&E&E&E&E&E\\
\hline
E&E&E&E&E&E&E\\
\hline
\end{array}
\end{eqnarray*}

If the non-erased locations of $W$ are zero, by Theorem~\ref{theo2}, the
array will be decoded as the zero array. Consider the array $V$
which has an
extra erasure in location $(u,v)=(0,1)$ (hence, $i= 0$ and
$j=3$ in
the proof of Theorem~\ref{cor00}), rendering

\begin{eqnarray*}
V&=&
\begin{array}{|c|c|c|c|c|c|c|}
\hline
\phantom{X}&E&\phantom{X}&\phantom{X}&\phantom{X}&\phantom{X}&E\\
\hline
\phantom{X}&\phantom{X}&\phantom{X}&\phantom{X}&\phantom{X}&\phantom{X}&E\\
\hline
\phantom{X}&\phantom{X}&\phantom{X}&\phantom{X}&E&E&E\\
\hline
\phantom{X}&\phantom{X}&\phantom{X}&E&E&E&E\\
\hline
E&E&E&E&E&E&E\\
\hline
E&E&E&E&E&E&E\\
\hline
\end{array}
\end{eqnarray*}

Consider the following arrays $Y^{(u',1)}$, $0\leq u'\leq 3$, defined
as in Theorem~\ref{cor00}, whose non-zero entries are denoted
$y^{(u',1)}_{a,b}$ below:

\begin{eqnarray*}
Y^{(0,1)}&=&
\begin{array}{|c|c|c|c|c|c|c|}
\hline
\;0\;&y^{(0,1)}_{0,1}&\;0\;&\;0\;&\;0\;&\;0\;&y^{(0,1)}_{0,6}\\
\hline
\;0\;&y^{(0,1)}_{1,1}&\;0\;&\;0\;&\;0\;&\;0\;&y^{(0,1)}_{1,6}\\
\hline
\;0\;&y^{(0,1)}_{2,1}&\;0\;&\;0\;&\;0\;&\;0\;&y^{(0,1)}_{2,6}\\
\hline
\;0\;&y^{(0,1)}_{3,1}&\;0\;&\;0\;&\;0\;&\;0\;&y^{(0,1)}_{3,6}\\
\hline
\;0\;&y^{(0,1)}_{4,1}&\;0\;&\;0\;&\;0\;&\;0\;&y^{(0,1)}_{4,6}\\
\hline
0&0&0&0&0&0&0\\
\hline
\end{array}
\end{eqnarray*}

\begin{eqnarray*}
Y^{(1,1)}&=&
\begin{array}{|c|c|c|c|c|c|c|}
\hline
0&0&0&0&0&0&0\\
\hline
\;0\;&y^{(1,1)}_{1,1}&\;0\;&\;0\;&\;0\;&\;0\;&y^{(1,1)}_{1,6}\\
\hline
\;0\;&y^{(1,1)}_{2,1}&\;0\;&\;0\;&\;0\;&\;0\;&y^{(1,1)}_{2,6}\\
\hline
\;0\;&y^{(1,1)}_{3,1}&\;0\;&\;0\;&\;0\;&\;0\;&y^{(1,1)}_{3,6}\\
\hline
\;0\;&y^{(1,1)}_{4,1}&\;0\;&\;0\;&\;0\;&\;0\;&y^{(1,1)}_{4,6}\\
\hline
\;0\;&y^{(1,1)}_{5,1}&\;0\;&\;0\;&\;0\;&\;0\;&y^{(1,1)}_{5,6}\\
\hline
\end{array}
\end{eqnarray*}

\begin{eqnarray*}
Y^{(2,1)}&=&
\begin{array}{|c|c|c|c|c|c|c|}
\hline
0&0&0&0&0&0&0\\
\hline
0&0&0&0&0&0&0\\
\hline
\;0\;&y^{(2,1)}_{2,1}&\;0\;&\;0\;&y^{(2,1)}_{2,4}&y^{(2,1)}_{2,5}&y^{(2,1)}_{2,6}\\
\hline
\;0\;&y^{(2,1)}_{3,1}&\;0\;&\;0\;&y^{(2,1)}_{3,4}&y^{(2,1)}_{3,5}&y^{(2,1)}_{3,6}\\
\hline
\;0\;&y^{(2,1)}_{4,1}&\;0\;&\;0\;&y^{(2,1)}_{4,4}&y^{(2,1)}_{4,5}&y^{(2,1)}_{4,6}\\
\hline
\;0\;&y^{(2,1)}_{5,1}&\;0\;&\;0\;&y^{(2,1)}_{5,4}&y^{(2,1)}_{5,5}&y^{(2,1)}_{5,6}\\
\hline
\end{array}
\end{eqnarray*}

\begin{eqnarray*}
Y^{(3,1)}&=&
\begin{array}{|c|c|c|c|c|c|c|}
\hline
0&0&0&0&0&0&0\\
\hline
0&0&0&0&0&0&0\\
\hline
0&0&0&0&0&0&0\\
\hline
\;0\;&y^{(3,1)}_{3,1}&\;0\;&y^{(3,1)}_{3,2}&y^{(3,1)}_{3,4}&y^{(3,1)}_{3,5}&y^{(3,1)}_{3,6}\\
\hline
\;0\;&y^{(3,1)}_{4,1}&\;0\;&y^{(3,1)}_{4,2}&y^{(3,1)}_{4,4}&y^{(3,1)}_{4,5}&y^{(3,1)}_{4,6}\\
\hline
\;0\;&y^{(3,1)}_{5,1}&\;0\;&y^{(3,1)}_{5,2}&y^{(3,1)}_{5,4}&y^{(3,1)}_{5,5}&y^{(3,1)}_{5,6}\\
\hline
\end{array}
\end{eqnarray*}

Such arrays with non-zero entries exist by Lemma~\ref{lemma1} (see
also Example~\ref{ex9}).
We choose $c_1$, $c_2$ and $c_3$  
such that

\begin{eqnarray*}
y^{(0,1)}_{1,1}\xor c_1y^{(1,1)}_{1,1}&=&0\\
y^{(0,1)}_{2,1}\xor c_1y^{(1,1)}_{2,1}\xor c_2y^{(2,1)}_{2,1}&=&0\\
y^{(0,1)}_{3,1}\xor c_1y^{(1,1)}_{3,1}\xor c_2y^{(2,1)}_{3,1}\xor c_3y^{(3,1)}_{3,1}&=&0\\
\end{eqnarray*}

Then, defining $Y=Y^{(0,1)}\xor c_1Y^{(1,1)}\xor c_2Y^{(2,1)}\xor
c_3Y^{(3,1)}$, we see that

\begin{eqnarray*}
Y&=&
\begin{array}{|c|c|c|c|c|c|c|}
\hline
\;0\;&y^{(0,1)}_{0,1}&\;0\;&\;0\;&\;0\;&\;0\;&X\\
\hline
0&0&0&0&0&0&X\\
\hline
0&0&0&0&X&X&X\\
\hline
0&0&0&X&X&X&X\\
\hline
0&X&0&X&X&X&X\\
\hline
0&X&0&X&X&X&X\\
\hline
\end{array}
\end{eqnarray*}
where entries denoted by $X$ may take any value.
Array $Y$ is non-zero since $y^{(0,1)}_{0,1}\neq 0$ . Array $V$
may be decoded as the zero array or as $Y$, so it is
uncorrectable. We can make the same argument for any entry
$(u,v)$ not contained in the erasures of $W$, so, by Lemma~\ref{lemma6},
the number of parity
symbols is 23 and the dimension of the code is 19.
}
\end{example}

The encoding is a special case of the decoding. For example, we may
place the parities at the end of the array in increasing order of
parities, as shown in Theorem~\ref{cor00} and in Example~\ref{ex10}.
The parities are considered as erasures and may be obtained using the
triangulation
method described in Theorem~\ref{theo2}. The fact that the locations
of the erasures
are known allows for a simplification of the decoding algorithm. For
example, the triangulated matrix corresponding to the coefficients
of~(\ref{eqGP11t}) may be precomputed. The next example illustrates
this encoding process.

\begin{example}
\label{ex77}
{\em
Take the 3-level EII code $\C(7,\hspace{-.2mm}(1,\hspace{-.2mm}1,
\hspace{-.2mm}3,\hspace{-.2mm}4,\hspace{-.2mm}7,\hspace{-.2mm}7)\hspace{-.2mm})$ of
Examples~\ref{ex7}, \ref{ex9} and~\ref{ex10} .
We have to solve the erasures in array $W$ of Example~\ref{ex10}
proceeding by triangulation like in Example~\ref{ex7}.

The first step is encoding rows $\uc_0$ and
$\uc_1$ (single parity). An ordering of the remaining rows in non-increasing number
of erasures is $\{i_0,i_1,i_2,i_3\}=\{5,4,3,2\}$.

According to~(\ref{eqGP2e3}), (\ref{eqGP1e3}) and~(\ref{eqGP1e3b}), and
rearranging the $\uc_i$s in non-increasing number
of erasures, we obtain

$$
\begin{array}{ccccccccccccl}
\uc_5&\hspace{-2mm}\xor\hspace{-2mm}
&\uc_4&\hspace{-2mm}\xor\hspace{-2mm} &
\uc_3&\hspace{-2mm}\xor\hspace{-2mm}
&\uc_2&\hspace{-2mm}\xor\hspace{-2mm}  &\uc_1&\hspace{-2mm}\xor\hspace{-2mm} &\uc_0&=&0\\
\al^{5}\uc_5&\hspace{-2mm}\xor\hspace{-2mm}
&\al^4\uc_4&\hspace{-2mm}\xor\hspace{-2mm}
&\al^3\uc_3&\hspace{-2mm}\xor\hspace{-2mm}  &\al^2\uc_2&\hspace{-2mm}\xor\hspace{-2mm}
&\al \uc_1&\hspace{-2mm}\xor\hspace{-2mm} &\uc_0&= &0\\
\al^{10}\uc_5&\hspace{-2mm}\xor\hspace{-2mm}
&\al^8\uc_4&\hspace{-2mm}\xor\hspace{-2mm}
&\al^6\uc_3&\hspace{-2mm}\xor\hspace{-2mm}  &\al^4\uc_2&\hspace{-2mm}\xor\hspace{-2mm}
&\al^2\uc_1&\hspace{-2mm}\xor\hspace{-2mm} &\uc_0&\in &\C_2\\
\al^{15}\uc_5&\hspace{-2mm}\xor\hspace{-2mm}
&\al^{12}\uc_4&\hspace{-2mm}\xor\hspace{-2mm}
&\al^9\uc_3&\hspace{-2mm}\xor\hspace{-2mm}  &\al^6\uc_2&\hspace{-2mm}\xor\hspace{-2mm}
&\al^3\uc_1&\hspace{-2mm}\xor\hspace{-2mm} &\uc_0&\in &\C_1\\\\
\end{array}
$$
Triangulating this linear system in $GF(8)$, 
since $\C_2\subset\C_1$ and $1\xor\al\xor\al^3=0$,
we obtain the
following triangulated system:

$$
\begin{array}{ccccccccccccl}
\uc_5&\hspace{-2mm}\xor\hspace{-2mm}
&\uc_4&\hspace{-2mm}\xor\hspace{-2mm}
&\uc_3&\hspace{-2mm}\xor\hspace{-2mm}
&\uc_2&\hspace{-2mm}\xor\hspace{-2mm}  &\uc_1&\hspace{-2mm}\xor\hspace{-2mm} &\uc_0&= &0\\
&&\uc_4&\hspace{-2mm}\xor\hspace{-2mm}
&\al^2\uc_3&\hspace{-2mm}\xor\hspace{-2mm}
&\al^3\uc_2&\hspace{-2mm}\xor\hspace{-2mm}  &\al^6 \uc_1&\hspace{-2mm}\xor\hspace{-2mm} &\al^4\uc_0&=&0\\
&&&&\uc_3&\hspace{-2mm}\xor\hspace{-2mm} &\al^3
\uc_2&\hspace{-2mm}\xor\hspace{-2mm}  &\uc_1&\hspace{-2mm}\xor\hspace{-2mm} &\al \uc_0&\in &\C_2\\
&&&&&&\uc_2&\hspace{-2mm}\xor\hspace{-2mm}
&\al^6\uc_1&\hspace{-2mm}\xor\hspace{-2mm}  &\al\uc_0&\in &\C_1.\\
\end{array}
$$

This triangulated system is precomputed, so it is not necessary to do
Gaussian elimination when encoding.

Then, $\uc_2\xor \al^6\uc_1\xor \al\uc_0$ is encoded in $\C_1$ and
$\uc_2$ is obtained as $\uc_2\eq (\uc_2\xor \al^6\uc_1\xor
\al\uc_0)\xor (\al^6\uc_1\xor \al\uc_0)$. Similarly,\\
$\uc_3\xor \al^3 \uc_2\xor \uc_1\xor \al \uc_0$ is encoded in $\C_2$,
and $\uc_3$ is obtained as $\uc_3\eq (\uc_3\xor \al^3 \uc_2\xor
\uc_1\xor \al \uc_0)\xor (\al^3 \uc_2\xor \uc_1\xor \al \uc_0)$.
Finally, we obtain
\begin{eqnarray*}
\uc_4&= &\al^2\uc_3\xor \al^3\uc_2\xor \al^6 \uc_1\xor \al^4\uc_0\\
\uc_5&= &\uc_4\xor \uc_3\xor \uc_2\xor \uc_1\xor \uc_0.
\end{eqnarray*}

At every step, we are encoding RS codes.
}
\end{example}

Another possibility for encoding EII codes is to use an existing encoding
algorithm for II codes. In effect, if $\uu$ is given by~(\ref{equu})
and $\uu'\hspace{-1mm}=\hspace{-1mm}
\left(\overbrace{u_0,\ldots,u_0}^{s_0},\overbrace{u_1,\ldots,u_1}^{s_1},\ldots,
\overbrace{u_{t-1},\ldots,u_{t-1}}^{s_{t-1}+s_t}\right)$, 
then an EII code $\C(n,\uu)$, in particular, is contained in an II code
$\C(n,\uu')$, both codes as given by Definition~\ref{defGPMDS}. Then,
by again setting the parities in the locations
described in Theorem~\ref{cor00} and in Example~\ref{ex10}, we 
obtain the vertical parities in locations $(a,b)$, $m-s_t\leq
a\leq m-1$, $0\leq b\leq n-u_{t-1}-1$ using~(\ref{eqGP1}) with $i\eq 0$. The
remaining parities are computed by encoding the data and the vertical
parities obtained in the previous step into the II code $\C(n,\uu')$.
Any encoding algorithm for II codes can be used, like, for example, the one
described in~\cite{w}.

The following theorem extends Theorem~II.2 on $t$-level
II codes as stated in~\cite{tk} and proven as Corollary~3
in~\cite{bh} (see also~\cite{w}).
It also
generalizes the well known result that the minimum distance of a
product code is the product of the minimum distances of the two
component codes.

\begin{theorem}
\label{cor11}
{\em
Consider the $t$-level EII code $\C(n,\uu)$ as given
by~Definition~\ref{defGPMDS}. Then,
the minimum distance of $\C(n,\uu)$ is

\begin{eqnarray}
\label{dist}
d&\hspace{-2mm} =\hspace{-2mm}&\min\left\{\left(\hs_{i+1}+1\right)\left(u_i+1\right)\;,\;0\leq
i\leq t-1\right\} 
\end{eqnarray}
}
\end{theorem}

\noindent\pf
For each $i$ such that $0\leq i\leq t-1$, consider an array in $\C(n,\uu)$
that has $\hs_{i+1}$ rows with $u_i+1$ erasures each, one row
with $u_i$ erasures, and all the other entries are zero. By
Theorem~\ref{theo2}, such arrays will be corrected by
the code $\C(n,\uu)$ as the zero codeword, thus

\begin{eqnarray*}
d&\geq &\min\left\{\left(\hs_{i+1}+1\right)\left(u_i+1\right)\;,\;0\leq
i\leq t-1\right\}.
\end{eqnarray*}

On the other hand, by Lemma~\ref{lemma1}, for each $0\leq i\leq t-1$,
there is an array in $\C(n,\uu)$ of weight
$\left(\hs_{i+1}+1\right)\left(u_i+1\right)$, so

\begin{eqnarray*}
d&\leq &\min\left\{\left(\hs_{i+1}+1\right)\left(u_i+1\right)\;,\;0\leq
i\leq t-1\right\}
\end{eqnarray*}
and~(\ref{dist}) follows.
\qed

\begin{example}
\label{ex8}
{\em
Consider the 3-level EII code $\C(7,\hspace{-.3mm}(1,\hspace{-.3mm}1,
\hspace{-.3mm}3,\hspace{-.3mm}4,\hspace{-.3mm}7,\hspace{-.3mm}7)\hspace{-.3mm})$ of
Examples~\ref{ex7}, \ref{ex9}, \ref{ex10} and~\ref{ex77}.
According to Theorem~\ref{cor11}, since $m=6$, $u_0=1$,
$u_1=3$, $u_2=4$, $s_0=2$, $s_1=1$, $s_2=1$ and $s_3=2$ (and
hence, $\hs_3=s_3=2$, $\hs_2=s_2+s_3=3$, $\hs_1=s_1+s_2+s_3=4$),
by~(\ref{dist}), the minimum distance of this code is
 $d=\min\left\{(5)(2)\,;(4)(4)\,;\,(3)(5)\right\}= 10$.
}
\end{example}

Although the minimum distance is not the only criterium to determine
the correction power of an II code~\cite{bh} (see also
Subsection~\ref{performance}), given different II 
codes as $m\times n$ arrays with the same rate, there is one that has
the largest minimum distance. A natural question is, if we include
EII codes, is there any EII code whose minimum distance is larger
than the one of any II code with the same rate? The answer depends on
the parameters chosen, but the following example shows that indeed
this may be the case.

\begin{example}
\label{ex88}
{\em
Consider the 4-level EII code $\C(7,(1,3,4,6,7))$.
According to Theorem~\ref{cor11}, the minimum distance of this code
is $d\eq 10$. An II code with the same rate is a code
$\C(7,(v_0,v_1,v_2,v_3,v_4))$, where $0\leq v_0\leq v_1\leq v_2\leq
v_3\leq v_4<7$ and $v_0+v_1+v_2+v_3+v_4\eq 21$. Again by
Theorem~\ref{cor11}, the minimum distance of this code
is $d\leq v_4+1\leq 7$, so the EII code has larger minimum distance than any II
code also consisting of $5\times 7$ arrays and with the same rate.
}
\end{example}

\subsection{Transpose Arrays, Iterative Decoding and Uniform
Distribution of Parity Symbols}
\label{transp}

Definition~\ref{defGPMDS} states that a $t$-level EII code $\C(n,\uu)$
consists of $m\times n$ arrays such that each row in the array is in
a code $\C_0$, and that certain linear combinations of the rows
belong in nested codes $\C_i$. If we take the columns in an array in $\C(n,\uu)$,
they would be the rows in an $n\times m$ transpose array. A natural
question is, are the rows in these transpose arrays also related by
certain nested codes?

Before answering this question,
we consider the simple example of a product code such that the vertical code is an
$[m,k_0,m-k_0+1]$ code and the horizontal code is an
$[n,k_1,n-k_1+1]$ code. We have seen in Example~\ref{ex4} that this
product code is a 1-level EII code $\C(n,\uu)$
where $\uu$ is the vector
$(\,\overbrace{n-k_1,n-k_1,\ldots,n-k_1}^{k_0},\overbrace{n,n,\ldots,n}^{m-k_0}\,)$.
If we consider the transpose arrays of the product code,
the rows of the transpose arrays (that is, the columns of the
original arrays) constitute a 1-level  EII code
$\C(m,\uu')$, where $\uu'$ is the vector
$(\,\overbrace{m-k_0,m-k_0,\ldots,m-k_0}^{k_1},\overbrace{m,m,\ldots,m}^{n-k_1}\,)$.
The following theorem generalizes this argument for
$t$-level EII codes.

\begin{theorem}
\label{theo3}
{\em
Consider a $t$-level EII code
$\C(n,\uu)$ as given by Definition~\ref{defGPMDS}, and take the set
of $n\times m$ transpose arrays corresponding 
to the $m\times n$ arrays in $\C(n,\uu)$. Then, this set of $n\times
m$ transpose arrays
constitute a $t$-level EII code $\C(m,\uu')$  such that, assuming $u_{-1}\eq 0$,

\begin{eqnarray}
\label{equu'}
\uu' &=&
\left(\overbrace{u'_0,u'_0,\ldots,u'_0}^{s'_0},\overbrace{u'_1,u'_1,\ldots,u'_1}^{s'_1},\ldots,
\overbrace{u'_{t-1},u'_{t-1},\ldots,u'_{t-1}}^{s'_{t-1}},\overbrace{u'_t,u'_t,\ldots,u'_t}^{s'_{t}}\right),
\end{eqnarray}
where
\begin{eqnarray}
\label{uprimeis}
u'_{t-i}=\hs_i
\hspace{-2mm}\quad {\rm and}\hspace{-2mm}\quad s'_i=
u_{t-i}-u_{t-i-1}\hspace{-2mm}\quad {\rm for}\hspace{-2mm}\quad 0\leq i\leq t.
\end{eqnarray}
}
\end{theorem}

\noindent\pf Denote by $\uc^{\rm\bf  (H)}_i$, $0\leq i\leq m-1$, the rows of an
array in $\C(n,\uu)$, and by $\uc^{\rm\bf  (V)}_j$, $0\leq j\leq
n-1$, the columns (that is, the rows of the $n\times m$ transpose array).
Specifically, if the array consists of symbols
$(c_{i,j})_{\substack{0\leq i\leq m-1\\ 0\leq j\leq n-1}}$, then
$\uc^{\rm\bf  (H)}_i=(c_{i,0},c_{i,1},\ldots,c_{i,n-1})$
for $0\leq i\leq m-1$ and $\uc^{\rm\bf  (V)}_j=(c_{0,j},c_{1,j},\ldots,c_{m-1,j})$
for $0\leq j\leq n-1$.

Consider the $t+1$ nested codes (on columns)
$\{0\}\eq\C'_t\subset\C'_{t-1}\subset\C'_{t-2}\subset\cdots\subset\C'_{0}$, where $\C'_i$
is an $[m,m-u'_i,u'_i+1]$ code and $u'_i$ is given by~(\ref{uprimeis}). A parity-check matrix of
$\C'_i$ is $H_{u'_i}$ as given by~(\ref{Hi}).

In order to prove the theorem, according to~(\ref{eqGP1}),
we have to prove that each
$\uc^{\rm\bf  (V)}_j\in\C'_{0}$ for $0\leq j\leq n-1$, and that

\begin{eqnarray}
\nonumber
\bigoplus_{j=0}^{n-1}\al^{rj}\uc^{\rm\bf  (V)}_j&\in&
\C'_{t-i}\;\;{\rm for}\;\;
0\leq i\leq t-1\\
\label{eqGP1'}
&& {\rm and}\;\; 0\leq r\leq \hat{s}'_{t-i}-1
\end{eqnarray}

$\C'_{0}$ is an $[m,m-u'_0,u'_0+1]$ code and by~(\ref{uprimeis}),
$u'_0=s_t$, so from~(\ref{eqGP1}), taking $i\eq 0$, $\uc^{\rm\bf  (V)}_j\in\C'_{0}$.

Next we have to
prove~(\ref{eqGP1'}). In effect, (\ref{eqGP1'}) holds
if and only if, by~(\ref{Hi}),

\begin{eqnarray*}
\bigoplus_{v=0}^{m-1}\al^{uv}\bigoplus_{j=0}^{n-1}\al^{rj}c_{v,j}&=&0
\end{eqnarray*}
for $0\leq i\leq t-1$, $0\leq u\leq u'_{t-i}-1$ and $0\leq r\leq \hat{s}'_{t-i}-1$,
if and only if, changing the summation order,

\begin{eqnarray*}
\bigoplus_{j=0}^{n-1}\al^{rj}\bigoplus_{v=0}^{m-1}\al^{uv}c_{v,j}&=&0
\end{eqnarray*}
for $0\leq i\leq t-1$, $0\leq u\leq u'_{t-i}-1$ and $0\leq r\leq \hat{s}'_{t-i}-1$,
if and only if, since, by~(\ref{hsi}) and~(\ref{uprimeis}),
$\hat{s}'_{t-i}=\sum_{z=t-i}^{t}s'_z=
\sum_{z=t-i}^{t}(u_{t-z}-u_{t-z-1})=u_{i}$ and
$u'_{t-i}=\hs_{i}$, by~(\ref{Hi}),

\begin{eqnarray*}
\bigoplus_{v=0}^{m-1}\al^{uv}\uc^{\rm\bf  (H)}_{v}&\hspace{-3mm}\in\hspace{-3mm} & \C_{i}\;\;{\rm for}\;\; 0\leq
i\leq t-1\;\;\hspace{-1mm}  {\rm and}\hspace{-1mm} \;\; 0\leq u\leq \hs_{i}-1,\\
\end{eqnarray*}
which is true by~(\ref{eqGP1}) and thus~(\ref{eqGP1'})
is also true.
\qed

Theorem~\ref{theo3} is the most important result in this section. One
application is an enhancement of the decoding algorithm by extending the
iterative decoding algorithm of product codes, in which rows
and columns are decoded iteratively until either all the erasures
are corrected or an uncorrectable pattern remains.
In order to illustrate this process, let us revisit
Example~\ref{ex7bis}.

\begin{example}
\label{ex7tris}
{\em
According to Theorem~\ref{theo3}, the transpose $7\times 4$ arrays
of the $4\times 7$ arrays in the 4-level II code $\C(7,(1,2,3,5))$
of Example~\ref{ex7bis} are in a 4-level EII code $\C(4,(0,0,1,1,2,3,4))$.
After (partially) decoding the rows of the array with erasures in
Example~\ref{ex7bis}, we were left with the uncorrectable array in $\C(7,(1,2,3,5))$

$$
\begin{array}{|c|c|c|c|c|c|c|}
\hline
E&\phantom{X}&\phantom{X}&E&\phantom{X}&E&E\\
\hline
&\phantom{X}&&\phantom{X}&&&\\
\hline
&&\phantom{X}&&&&\\
\hline
E&E&&&&E&E\\
\hline
\end{array}
$$

Notice that this array has two columns with no erasures, two columns
with one erasure each and three columns with two erasures each. By
Theorem~\ref{theo2}, the array is correctable in $\C(4,(0,0,1,1,2,3,4))$. Hence,
after two iterations the erasures are corrected.
}
\end{example}

\begin{example}
\label{ex7quatris}
{\em
Consider the 5-level II code $\C(10,(1,3,6,8,9))$. 
The transpose arrays of $\C(10,(1,3,6,8,9))$
are in the 5-level EII code $\C(5,(0,1,2,2,3,3,3,4,4,5))$
by Theorem~\ref{theo3}. Assume that
the following array is received:

$$
\begin{array}{|c|c|c|c|c|c|c|c|c|c|}
\hline
E&\phantom{X}&\phantom{X}&\phantom{X}&E&E&\phantom{X}&E&\phantom{X}&\phantom{X}\\
\hline
\phantom{X}&E&E&\phantom{X}&E&E&E&E&\phantom{X}&E\\
\hline
\phantom{X}&\phantom{X}&\phantom{X}&\phantom{X}&\phantom{X}&\phantom{X}&\phantom{X}&\phantom{X}&E&\phantom{X}\\
\hline
E&E&E&\phantom{X}&&E&E&E&E&E\\
\hline
E&E&E&\phantom{X}&\phantom{X}&E&E&E&\phantom{X}&E\\
\hline
\end{array}
$$

Applying the decoding algorithm on rows, only the third row
can be corrected, since it has exactly one erasure.
After correction of the third row, we have the array

$$
\begin{array}{|c|c|c|c|c|c|c|c|c|c|}
\hline
E&\phantom{X}&\phantom{X}&\phantom{X}&E&E&\phantom{X}&E&\phantom{X}&\phantom{X}\\
\hline
\phantom{X}&E&E&\phantom{X}&E&E&E&E&\phantom{X}&E\\
\hline
\phantom{X}&\phantom{X}&\phantom{X}&\phantom{X}&\phantom{X}&\phantom{X}&\phantom{X}&\phantom{X}&&\phantom{X}\\
\hline
E&E&E&\phantom{X}&&E&E&E&E&E\\
\hline
E&E&E&\phantom{X}&\phantom{X}&E&E&E&\phantom{X}&E\\
\hline
\end{array}
$$

This array has one column with no erasures, one column with one
erasure and one column with two erasures, while the remaining columns
contain more than two erasures. The decoding algorithm on columns
(i.e., on the code $\C(5,(0,1,2,2,3,3,3,4,4,5))$)
allows for correction of the column with one erasure and the column
with two erasures, giving the array

$$
\begin{array}{|c|c|c|c|c|c|c|c|c|c|}
\hline
E&\phantom{X}&\phantom{X}&\phantom{X}& &E&\phantom{X}&E&\phantom{X}&\phantom{X}\\
\hline
\phantom{X}&E&E&\phantom{X}& &E&E&E&\phantom{X}&E\\
\hline
\phantom{X}&\phantom{X}&\phantom{X}&\phantom{X}&\phantom{X}&\phantom{X}&\phantom{X}&\phantom{X}&&\phantom{X}\\
\hline
E&E&E&\phantom{X}&&E&E&E& &E\\
\hline
E&E&E&\phantom{X}&\phantom{X}&E&E&E&\phantom{X}&E\\
\hline
\end{array}
$$

This last array is decodable on rows (i.e., on the code
$\C(10,(1,3,6,8,9))$), hence, the erasures are corrected
after three iterations.
}
\end{example}

A second application of Theorem~\ref{theo3} is allowing for a
balanced distribution of the parity symbols in EII codes. In effect,
given an $[mn,k]$  
code consisting of $m\times n$ arrays, if $mn-k\eq qm+r$, where
$0\leq r<m$, we say that the code has a balanced distribution of
parity symbols if $m-r$ of the rows contain $q$ parity symbols, while
the remaining $r$ rows contain $q+1$ parity symbols.
Codes somewhat similar to II codes with a balanced distribution of parity symbols
were presented in~\cite{chpb}.  Actually, in~\cite{chpb}
only cases for which $r\eq 0$ are considered, i.e., $m$ divides $mn-k$ and
hence each row contains the same
number $q$ of parity symbols.

Given a $t$-level EII code $\C(n,\uu)$, so far
we have placed the parity symbols as in Theorem~\ref{cor00} and in examples~\ref{ex10}
and~\ref{ex77}, that is, at the end of each row in
non-decreasing order of the $u_i$s. However, this distribution of
symbols in general is not balanced. 
If it can be shown that there is an uniform distribution of
erasures that can be corrected
by the code $\C(m,\uu')$ (i.e., the code on columns as given by
Theorem~\ref{theo3}), then we can use those erasures as the locations 
for the parity symbols.
The following theorem shows that,
using Theorem~\ref{theo3}, we
can easily obtain a balanced distribution of the parity symbols for a
$t$-level EII code $\C(n,\uu)$.

\begin{theorem}
\label{theo4}
{\em
Consider a $t$-level EII code $\C(n,\uu)$ as given by
Definition~\ref{defGPMDS}. Then $\C(n,\uu)$ admits a balanced distribution
of the parity symbols.
}
\end{theorem}

\noindent\pf We need to find $s\eq\sum_{i=0}^ts_iu_i$ erasures such that,
if $s\eq qm+r$ with $0\leq r<m$, then there are
$m-r$ rows with $q$ erasures each and $r$ rows with $q+1$
erasures each, and the erasures are correctable by the the $t$-level
EII code $\C(m,\uu')$ on columns as given by 
Theorem~\ref{theo3}. Then such erasures can be used to place the
parity symbols. 

In effect, let $v_0\geq v_1\geq \ldots\geq v_{z-1}$ be the
non-zero elements of $\uu'$ in non-increasing order. In particular,
$s\eq \sum_{i=0}^{z-1}v_i$. We will
select the first $z$ columns in an $m\times n$ array such that column
$j$ has exactly $v_j$ erasures for $0\leq j\leq z-1$. Then, by
Theorem~\ref{theo3}, such erasures are correctable. In addition, we
will show that the selection of erasures is balanced. We proceed by
induction.

If $z\eq 1$, we have only one column and we place the erasures in the
top $v_0$ positions of that column. In particular, the distribution
is balanced. So assume that $z>1$.

Consider the first $z-1$ columns and let $s'\eq \sum_{i=0}^{z-2}v_i$. By
induction, if $s'\eq q'm+r'$, we can place $s_j$ erasures in column
$j$ for $0\leq j\leq z-2$, such that the first $r'$ rows contain
$q'+1$ erasures and the last $m-r'$ rows contain $q'$ erasures.

If $v_{z-1}\,\leq\,m-r'$, then in column $z-1$ we place the $v_{z-1}$
erasures in locations $r',r'+1,\ldots ,r'+v_{z-1}-1$. Then the first
$r'+v_{z-1}$ rows contain $q'+1$ erasures and the last
$m-(r'+v_{z-1})$ rows contain $q'$ erasures, giving a balanced
distribution of the erasures.

If $v_{z-1}>m-r'$, then in column $z-1$ we place the $v_{z-1}$
erasures in locations $$0,1,\ldots ,v_{z-1}-(m-r')-1,
r',r'+1,\ldots ,m-1.$$ Then the first $v_{z-1}-(m-r')$ rows contain
$q'+1$ erasures and the
remaining rows $q'$ erasures, also giving a
balanced distribution of the erasures.
\qed

Of course the balanced distribution of parity symbols is not unique.
We illustrate the method described
in Theorem~\ref{theo4} in the next two examples.

\begin{example}
\label{ex13}
{\em
Consider a product code consisting of $5\times 7$ arrays such that
each row has one parity and each column two parities. We have seen
in Example~\ref{ex4} that such a code can be viewed as a 1-level EII
code $\C(7,(1,1,1,7,7))$. The distribution of parities given in the
proof of Theorem~\ref{theo4} in this case is the following:

$$
\begin{array}{|c|c|c|c|c|c|c|}
\hline
E&E&&E&\phantom{X}&\phantom{X}&E\\
\hline
E&E&&&E&&E\\
\hline
E&&E&&E&&\\
\hline
E&&E&&&E&\\
\hline
E&&&E&&E&\\
\hline
\end{array}
$$

}
\end{example}

Certainly it is not necessary to invoke Theorem~\ref{theo4} to obtain
a balanced distribution of the parities in a product code. The next
example is more representative.

\begin{example}
\label{ex14}
{\em
Consider the 3-level EII code
$\C(7,\hspace{-.2mm}(1,\hspace{-.2mm}1,\hspace{-.2mm}3,\hspace{-.2mm}4,\hspace{-.2mm}7, 
\hspace{-.2mm}7)\hspace{-.2mm})$ of
Examples~\ref{ex9}, \ref{ex10}, \ref{ex77} and~\ref{ex8}.
According to Theorem~\ref{theo3}, the $7\times 6$ transpose arrays of
this code
constitute a 3-level EII code $\C(6,(2,2,2,3,4,4,6))$.
The balanced distribution of
parities given by Theorem~\ref{theo4} is

$$
\begin{array}{|c|c|c|c|c|c|c|}
\hline
E&E&E&&E&&\\
\hline
E&E&E&&&E&\\
\hline
E&E&&E&&E&\\
\hline
E&E&&E&&&E\\
\hline
E&&E&E&&&E\\
\hline
E&&E&&E&&\\
\hline
\end{array}
$$

}
\end{example}

For encoding using a balanced distribution of parities, we apply the
decoding algorithm by triangulation on the $t$-level EII code
$\C(m,\uu')$ on columns given by Theorem~\ref{theo3}. Again, the fact
that the erasures are known a priori allows for precomputing the
coefficients arising from the triangulation. Obviously, any $t$-level
EII code $\C(n,\uu)$ also admits a balanced distribution of symbols
on columns.

\subsection{Error and Erasure Decoding of EII Codes}
\label{errorerasure}

Although in this paper we concentrate on the erasure model, 
the decoding algorithm can be adapted to handle
errors together with erasures. Specifically: 

\begin{algorithm}
\label{errors}
{\em
Consider a $t$-level EII code $\C(n,\uu)$ as given by
Definition~\ref{defGPMDS} and assume that a received $m\times n$ array
contains both errors and erasures. Then proceed as follows:

\begin{enumerate}

\item Attempt to correct in $\C_0$ rows with
up to $i$ errors together with up to $j$
erasures, where $2i+j\leq u_0$.

\item Consider the $\ell$ rows for which the decoding has failed. If
$\ell\eq 0$, then correction has been successful and exit the
algorithm.

\item If $\ell>\hs_1$, then declare that the algorithm has
failed. Otherwise, as in Theorem~\ref{theo2},
let $i_0,i_1,\ldots,i_{m-1}$
be an ordering of the rows according to a
non-increasing number of erasures such that rows
$i_0,i_1,\ldots,i_{\ell -1}$ correspond to the $\ell$ rows for which
the decoding in $\C_0$ has failed.

\item Define $w$ as in~(\ref{w}), i.e., $\hs_{w+1}<\ell\leq
\hs_w$ and consider the
code $\C_w$ from the nested set of codes in
Definition~\ref{defGPMDS}, which can correct up to $i$ errors
together with up to $j$ erasures for $2i+j\leq u_w$.

\item Proceeding as in Theorem~\ref{theo2}, after triangulation,
obtain~(\ref{eqGP11tl}). Then attempt to correct up to $i$ errors
together with up to $j$ erasures, where $2i+j\leq u_w$. If the
decoding is successful, continue by induction with the remaining
$\ell -1$ rows. If the decoding is unsuccessful, change the order
of the $\ell$ uncorrected rows (for example, by rotating them) and repeat the
procedure until $\uc_{i_{\ell -1}}$ is decoded successfully and then
proceed by induction. If none of the $\ell$ rows is decoded
successfully after this procedure, declare failure.

\end{enumerate}

If the algorithm fails, then correction is attempted on columns. \qed
}
\end{algorithm}

Contrary to the case of erasures only, there is now a
probability of miscorrection each time individual decoding or errors
together with erasures is attempted: if the error-erasure correcting power
of the codes is exceeded, the decoder may miscorrect and give the wrong codeword.
However, if the finite field is fairly large and the codes can
correct a substantial number of errors, such probability is
small~\cite{kmc,kmc2,msw} and we assume that miscorrection does not
occur (otherwise, the decoding algorithm gets more complicated). A
similar assumption was made in~\cite{w}. Certainly, also the decoding
algorithms of~\cite{tk} and~\cite{w} can be adapted for $t$-level EII
codes. 

We illustrate Algorithm~\ref{errors} with an example.

\begin{example}
\label{ex15}
{\em
Consider $\C(15,(3,3,5,8,8,15))$ as a 3-level EII code  over the
field $GF(16)$, hence, $\C_0$ is a $[15,12,4]$ code, $\C_1$ is a
$[15,10,6]$ code, $\C_2$ is a $[15,7,9]$ code and $\C_3\eq\{0\}$.
Denoting errors by $X$ and erasures by $E$, assume that the following
array has been received:

$$
\begin{array}{c|c|c|c|c|c|c|c|c|c|c|c|c|c|c|c|}
\cline{2-16}
\uc_0&\phantom{\hspace{-.2mm}X\hspace{-.2mm}}&\hspace{-.2mm}X\hspace{-.2mm}&
\phantom{\hspace{-.2mm}X\hspace{-.2mm}}&\phantom{\hspace{-.2mm}X\hspace{-.2mm}}&\phantom{\hspace{-.2mm}X\hspace{-.2mm}}&
\hspace{-.2mm}X\hspace{-.2mm}&\phantom{\hspace{-.2mm}X\hspace{-.2mm}}&
\phantom{\hspace{-.2mm}X\hspace{-.2mm}}&\hspace{-.2mm}X\hspace{-.2mm}&
\phantom{\hspace{-.2mm}X\hspace{-.2mm}}&
\phantom{\hspace{-.2mm}X\hspace{-.2mm}}&\hspace{-.2mm}X\hspace{-.2mm}&
\phantom{\hspace{-.2mm}X\hspace{-.2mm}}&\phantom{\hspace{-.2mm}X\hspace{-.2mm}}&\phantom{\hspace{-.2mm}X\hspace{-.2mm}}\\
\cline{2-16}
\uc_1&\phantom{\hspace{-.2mm}X\hspace{-.2mm}}&\phantom{\hspace{-.2mm}X\hspace{-.2mm}}&
\phantom{\hspace{-.2mm}X\hspace{-.2mm}}&\phantom{\hspace{-.2mm}X\hspace{-.2mm}}&\hspace{-.2mm}X\hspace{-.2mm}&
\phantom{\hspace{-.2mm}X\hspace{-.2mm}}&\phantom{\hspace{-.2mm}X\hspace{-.2mm}}&
\phantom{\hspace{-.2mm}X\hspace{-.2mm}}&E&\phantom{\hspace{-.2mm}X\hspace{-.2mm}}&
\phantom{\hspace{-.2mm}X\hspace{-.2mm}}&\phantom{\hspace{-.2mm}X\hspace{-.2mm}}&
\phantom{\hspace{-.2mm}X\hspace{-.2mm}}&\phantom{\hspace{-.2mm}X\hspace{-.2mm}}&\phantom{\hspace{-.2mm}X\hspace{-.2mm}}\\
\cline{2-16}
\uc_2&\hspace{-.2mm}X\hspace{-.2mm}&\phantom{\hspace{-.2mm}X\hspace{-.2mm}}&
\phantom{\hspace{-.2mm}X\hspace{-.2mm}}&\hspace{-.2mm}X\hspace{-.2mm}&
\phantom{\hspace{-.2mm}X\hspace{-.2mm}}&\phantom{\hspace{-.2mm}X\hspace{-.2mm}}&
\hspace{-.2mm}X\hspace{-.2mm}&\phantom{\hspace{-.2mm}X\hspace{-.2mm}}&
\phantom{\hspace{-.2mm}X\hspace{-.2mm}}&\phantom{\hspace{-.2mm}X\hspace{-.2mm}}&\phantom{\hspace{-.2mm}X\hspace{-.2mm}}&
\hspace{-.2mm}X\hspace{-.2mm}&\phantom{\hspace{-.2mm}X\hspace{-.2mm}}&
\hspace{-.2mm}X\hspace{-.2mm}&\phantom{\hspace{-.2mm}X\hspace{-.2mm}}\\
\cline{2-16}
\uc_3&\phantom{\hspace{-.2mm}X\hspace{-.2mm}}&\phantom{\hspace{-.2mm}X\hspace{-.2mm}}&
\phantom{\hspace{-.2mm}X\hspace{-.2mm}}&\phantom{\hspace{-.2mm}X\hspace{-.2mm}}&\phantom{\hspace{-.2mm}X\hspace{-.2mm}}&
\hspace{-.2mm}X\hspace{-.2mm}&\phantom{\hspace{-.2mm}X\hspace{-.2mm}}&\phantom{\hspace{-.2mm}X\hspace{-.2mm}}
&\phantom{\hspace{-.2mm}X\hspace{-.2mm}}&\phantom{\hspace{-.2mm}X\hspace{-.2mm}}&
\hspace{-.2mm}X\hspace{-.2mm}&\phantom{\hspace{-.2mm}X\hspace{-.2mm}}&\phantom{\hspace{-.2mm}X\hspace{-.2mm}}
&\phantom{\hspace{-.2mm}X\hspace{-.2mm}}&\phantom{\hspace{-.2mm}X\hspace{-.2mm}}\\
\cline{2-16}
\uc_4&\phantom{\hspace{-.2mm}X\hspace{-.2mm}}&\hspace{-.2mm}X\hspace{-.2mm}&\phantom{\hspace{-.2mm}X\hspace{-.2mm}}
&\phantom{\hspace{-.2mm}X\hspace{-.2mm}}&\phantom{\hspace{-.2mm}X\hspace{-.2mm}}&
\phantom{\hspace{-.2mm}X\hspace{-.2mm}}&\phantom{\hspace{-.2mm}X\hspace{-.2mm}}&
\phantom{\hspace{-.2mm}X\hspace{-.2mm}}&\phantom{\hspace{-.2mm}X\hspace{-.2mm}}&\phantom{\hspace{-.2mm}X\hspace{-.2mm}}&
\phantom{\hspace{-.2mm}X\hspace{-.2mm}}&E&\phantom{\hspace{-.2mm}X\hspace{-.2mm}}&
\phantom{\hspace{-.2mm}X\hspace{-.2mm}}&\phantom{\hspace{-.2mm}X\hspace{-.2mm}}\\
\cline{2-16}
\uc_5&E&\phantom{\hspace{-.2mm}X\hspace{-.2mm}}&\hspace{-.2mm}X\hspace{-.2mm}&\phantom{\hspace{-.2mm}X\hspace{-.2mm}}&E&
\hspace{-.2mm}X\hspace{-.2mm}&\phantom{\hspace{-.2mm}X\hspace{-.2mm}}&
\phantom{\hspace{-.2mm}X\hspace{-.2mm}}&\phantom{\hspace{-.2mm}X\hspace{-.2mm}}&\phantom{\hspace{-.2mm}X\hspace{-.2mm}}&
E&\phantom{\hspace{-.2mm}X\hspace{-.2mm}}&\phantom{\hspace{-.2mm}X\hspace{-.2mm}}&E&\phantom{\hspace{-.2mm}X\hspace{-.2mm}}\\
\cline{2-16}
\end{array}
$$

Following Algorithm~\ref{errors}, correction of up to one error
together with one erasure is attempted using code $\C_0$. This
correction succeeds for rows $\uc_1$ and $\uc_4$, and fails in the
remaining 4 rows (i.e., $\ell\eq 4$). Ordering the rows in
non-decreasing number of erasures and according to~(\ref{eqGP2e3}),
(\ref{eqGP1e3}) and~(\ref{eqGP1e3b}) as in Example~\ref{ex7}, we
obtain (now rows $\uc_1$ and $\uc_4$ are error-free)

$$
\begin{array}{ccccccccccccl}
\uc_5&\hspace{-2mm}\xor\hspace{-2mm}&\uc_0&\hspace{-2mm}\xor\hspace{-2mm}&\uc_2&
\hspace{-2mm}\xor\hspace{-2mm}&\uc_3&\hspace{-2mm}\xor\hspace{-2mm}&\uc_1&\hspace{-2mm}\xor\hspace{-2mm}&\uc_4&= &0\\
\al^5\uc_5&\hspace{-2mm}\xor\hspace{-2mm}&\uc_0&\hspace{-2mm}\xor\hspace{-2mm}&
\al^2\uc_2&\hspace{-2mm}\xor\hspace{-2mm}&\al^3\uc_3&\hspace{-2mm}\xor\hspace{-2mm}&\al\uc_1&
\hspace{-2mm}\xor\hspace{-2mm}&\al^4\uc_4&\in &\C_2\\
\al^{10}\uc_5&\hspace{-2mm}\xor\hspace{-2mm}&\uc_0&\hspace{-2mm}\xor\hspace{-2mm}&\al^4\uc_2&
\hspace{-2mm}\xor\hspace{-2mm}&\al^6\uc_3&\xor
&\al^2\uc_1&\hspace{-2mm}\xor\hspace{-2mm}&\al^8\uc_4&\in &\C_2\\
\al^{15}\uc_5&\hspace{-2mm}\xor\hspace{-2mm}&\uc_0&\hspace{-2mm}\xor\hspace{-2mm}&\al^6\uc_2&
\hspace{-2mm}\xor\hspace{-2mm}&\al^9\uc_3&\hspace{-2mm}\xor\hspace{-2mm}&
\al^3\uc_1&\hspace{-2mm}\xor\hspace{-2mm}&\al^{12}\uc_4&\in &\C_1.\\
\end{array}
$$

Triangulating this linear system in $GF(16)$ and assuming that $1\xor\al\xor\al^4=0$,
we obtain 

$$
\begin{array}{ccccccccccccl}
\uc_5&\hspace{-2mm}\xor\hspace{-2mm}&\uc_0&\hspace{-2mm}\xor\hspace{-2mm}&\uc_2&
\hspace{-2mm}\xor\hspace{-2mm}&\uc_3&\hspace{-2mm}\xor\hspace{-2mm}&\uc_1&\hspace{-2mm}\xor\hspace{-2mm}&\uc_4&= &0\\
&&\uc_0&\hspace{-2mm}\xor\hspace{-2mm}&\al^6\uc_2&\hspace{-2mm}\xor\hspace{-2mm}&
\al\uc_3&\hspace{-2mm}\xor\hspace{-2mm}&\al^7\uc_1&\hspace{-2mm}\xor\hspace{-2mm}&\al^{13}\uc_4&\in &\C_2\\
&&&&\uc_2&\hspace{-2mm}\xor\hspace{-2mm}&\al\uc_3&\hspace{-2mm}\xor\hspace{-2mm}&\al^{12}\uc_1&
\hspace{-2mm}\xor\hspace{-2mm}&\uc_4&\in &\C_2\\
&&&&&&\uc_3&\hspace{-2mm}\xor\hspace{-2mm}&\al^{10}\uc_1&\hspace{-2mm}\xor\hspace{-2mm}&\al^3\uc_4&\in &\C_1.\\
\end{array}
$$

Since $\uc_3\xor \al^{10}\uc_1\xor \al^3\uc_4$ has two errors, they
can be corrected in $\C_1$. So, we get
$$\uc_3\;\eq\; \left(\uc_3\xor \al^{10}\uc_1\xor \al^3\uc_4\right)\xor
\left(\al^{10}\uc_1\xor \al^3\uc_4\right).$$

Next we attempt to decode $\uc_2\xor\al\uc_3\xor \al^{12}\uc_1\xor
\uc_4$ in $\C_2$. But this vector has five errors, which are
uncorrectable in $\C_2$. Making a rotation of the 3 uncorrected
vectors, we have

$$
\begin{array}{ccccccccccccl}
\uc_2&\hspace{-2mm}\xor\hspace{-2mm}&\uc_5&\hspace{-2mm}\xor\hspace{-2mm}&\uc_0&
\hspace{-2mm}\xor\hspace{-2mm}&\uc_3&\hspace{-2mm}\xor\hspace{-2mm}&\uc_1&\hspace{-2mm}\xor\hspace{-2mm}&\uc_4&= &0\\
\al^2\uc_2&\hspace{-2mm}\xor\hspace{-2mm}&\al^5\uc_5&\hspace{-2mm}\xor\hspace{-2mm}&\uc_0 
&\hspace{-2mm}\xor\hspace{-2mm}&\al^3\uc_3&\hspace{-2mm}\xor\hspace{-2mm}&\al\uc_1&
\hspace{-2mm}\xor\hspace{-2mm}&\al^4\uc_4&\in &\C_2\\
\al^4\uc_2&\hspace{-2mm}\xor\hspace{-2mm}&\al^{10}\uc_5&\hspace{-2mm}\xor\hspace{-2mm}&\uc_0&
\hspace{-2mm}\xor\hspace{-2mm}&\al^6\uc_3&\xor
&\al^2\uc_1&\hspace{-2mm}\xor\hspace{-2mm}&\al^8\uc_4&\in &\C_2.\\
\end{array}
$$

Triangulating this linear system, we obtain

$$
\begin{array}{ccccccccccccl}
\uc_2&\hspace{-2mm}\xor\hspace{-2mm}&\uc_5&\hspace{-2mm}\xor\hspace{-2mm}&\uc_0&
\hspace{-2mm}\xor\hspace{-2mm}&\uc_3&\hspace{-2mm}\xor\hspace{-2mm}&\uc_1&\hspace{-2mm}\xor\hspace{-2mm}&\uc_4&= &0\\
&&\uc_5&\hspace{-2mm}\xor\hspace{-2mm}&\al^7\uc_0&\hspace{-2mm}\xor\hspace{-2mm}&\al^5\uc_3&
\hspace{-2mm}\xor\hspace{-2mm}&\al^4\uc_1&\hspace{-2mm}\xor\hspace{-2mm}&\al^{9}\uc_4&\in &\C_2\\
&&&&\uc_0&\hspace{-2mm}\xor\hspace{-2mm}&\al^{14}\uc_3&\hspace{-2mm}\xor\hspace{-2mm}&\al^{4}\uc_1&
\hspace{-2mm}\xor\hspace{-2mm}&\uc_4&\in &\C_2.\\
\end{array}
$$

Now, $\uc_0\xor \al^{14}\uc_3\xor \al^{4}\uc_1\xor \uc_4$ has four
errors, that are correctable in $\C_2$, so $\uc_0$ is obtained as
\begin{eqnarray*}
\uc_0&=&\left(\uc_0\xor \al^{14}\uc_3\xor \al^{4}\uc_1\xor
\uc_4\right)\xor \left(\al^{14}\uc_3\xor \al^{4}\uc_1\xor
\uc_4\right).\end{eqnarray*}

Next, $\uc_5\xor \al^7\uc_0\xor \al^5\uc_3\xor \al^4\uc_1\xor
\al^{9}\uc_4$ has two errors and four erasures, which are also
correctable in $\C_2$, so

\begin{eqnarray*} 
\uc_5&=&\left(\uc_5\xor \al^7\uc_0\xor \al^5\uc_3\xor \al^4\uc_1\xor
\al^{9}\uc_4\right)\xor \left(\al^7\uc_0\xor \al^5\uc_3\xor \al^4\uc_1\xor
\al^{9}\uc_4\right).\end{eqnarray*}

Finally, 
\begin{eqnarray*}
\uc_2&\hspace{-3mm}=\hspace{-3mm}&(\uc_2\hspace{-.2mm}\xor\hspace{-.2mm}\uc_5\hspace{-.2mm}
\xor\hspace{-.2mm}\uc_0\hspace{-.2mm}\xor\hspace{-.2mm}\uc_3\xor
\uc_1\hspace{-.2mm}\xor\hspace{-.2mm}\uc_4)\hspace{-.2mm}\xor\hspace{-.2mm}
(\uc_5\hspace{-.2mm}\xor\hspace{-.2mm}\uc_0\hspace{-.2mm}\xor\hspace{-.2mm}\uc_3\hspace{-.2mm}
\xor\hspace{-.2mm}\uc_1\hspace{-.2mm}\xor\hspace{-.2mm}\uc_4),\end{eqnarray*}
completing the decoding.
}
\end{example}

\section{Extended Product Codes and Optimality Issues}
\label{optimality}
This section is structured as follows: in
Subsection~\ref{upperbound}, we present an upper bound on the minimum
distance of EPC codes, we illustrate it with examples and we show
that other bounds, like the bound on LRC codes, are special cases of
this bound. In Subsection~\ref{optimalEPC} we present some
constructions of codes meeting this upper bound for important special cases.
In Subsection~\ref{performance}, we briefly discuss tradeoffs between
different codes by giving Monte Carlo simulations for some specific parameters.

\subsection{Upper Bound on the Minimum Distance of Extended Product Codes}
\label{upperbound}
The $t$-level EII codes $\C(n,\uu)$ described in Section~\ref{GP}
are a special case of product codes with some extra
parities defined in Section~\ref{Introduction}, where
we called these codes extended product (EPC) codes and we denoted them
by $EP(m,v;n,h;g)$, $v$ the number of vertical
parities in each column, $h$ the number of horizontal parities in
each row, and $g$ the number
of extra parities.
From Definition~\ref{defGPMDS}, it is easy to determine $v$, $h$ and
$g$ for $t$-level EII codes $\C(n,\uu)$. In effect, since $v\eq
s_t$ and $h\eq u_0$, the extra parities consist of all the remaining
parities, i.e., $g\eq \left(\sum_{i=0}^tu_is_i\right)-u_0m-s_t(n-u_0)$.
For example, the 3-level EII code $\C(7,(1,1,3,4,7,7))$ of
Examples~\ref{ex7}, \ref{ex9}, \ref{ex10} and~\ref{ex14} is an $EP(6,2;7,1;5)$ code.

The next theorem gives an upper bound on the minimum distance of an
$EP(m,v;n,h;g)$ code.

\begin{theorem}
\label{lemma2}
{\em
Let $d(m,v;n,h;g)$ be the minimum distance of an $EP(m,v;n,h;g)$
code. Given $a$ such that $1\leq a\leq g+1$,
$b=\lfloor (g+1)/a\rfloor$ and $r=g+1-ab$, let

\begin{eqnarray}
\label{dvhga0}
\hspace{-8mm}d(v,h,g;a)&\hspace{-2mm}=\hspace{-2mm}&(v+b)(h+a)\quad {\rm if}\quad r=0\\
\label{dvhgar}
\hspace{-8mm}d(v,h,g;a)&\hspace{-2mm}=\hspace{-2mm}&(v+b)(h+a)+h+r\quad\hspace{-2mm} {\rm if}\hspace{-2mm}\quad r\,\neq\,0
\end{eqnarray}
Then,
\begin{eqnarray}
\label{dvhg}
d(m,v;n,h;g)&\hspace{-2mm}\leq\hspace{-2mm} &\min\{d(v,h,g\,;\,a)\,:\,\lceil (g+1)/(m-v)\rceil
\leq a\leq \min\{g+1\,,\,n-h\}\}
\end{eqnarray}
}
\end{theorem}

\noindent\pf Assume first that $r=0$, the zero array is stored, and the received
array has the $(v+b)(h+a)$ locations $(i,j)$ erased, where
$0\leq i\leq v+b-1$ and $0\leq j\leq h+a-1$. Notice that, since
$\lceil (g+1)/(m-v)\rceil\leq a\leq \min\{g+1\,,\,n-h\}$,
\begin{eqnarray*}
0\leq i\leq v+b-1\leq
v+{\frac{g+1}{(g+1)/(m-v)}}-1= m-1
\end{eqnarray*}
and $0\leq j\leq h+a-1\leq h+(n-h)-1=n-1$, so all the erasures
are within the array.
The erasures are covered by $h(v+b)$ horizontal
parities, $v(h+a)$ vertical parities and $g$ extra parities, but $hv$ of such
parities are dependent. Since $ab= g+1$, there are only
$(v+b)(h+a)-1$ independent parities covering the $(v+b)(h+a)$
erasures, insufficient to correct them.

Similarly, assume that $r\neq 0$, the zero array is stored, and the received
array has the $(v+b)(h+a)+h+r$ locations $(i,j)$ erased, where either
$0\leq i\leq v+b-1$ and $0\leq j\leq h+a-1\leq n-1$, or
$i= v+b$ and $0\leq j\leq h+r-1$.
Observe that all the erasures are within the array. In effect,
since $a$ does not divide $g+1$,
\begin{eqnarray*}
v+b&\hspace{-2mm}=\hspace{-2mm}&v+\left\lfloor{\frac{g+1}{a}}\right\rfloor<v+{\frac{g+1}{a}}
\leq v+{\frac{g+1}{\lceil(g+1)/(m-v)\rceil}}\,\leq\,v+(m-v)=m.
\end{eqnarray*}

The erasures are covered by $h(v+b+1)$ horizontal parities, $v(h+a)$ vertical parities
and $g$ extra parities. Since $hv$ of such parities are
dependent and $g+1=ab+r$, this gives a total of
$(v+b)(h+a)+h+r-1$ independent parities covering the $(v+b)(h+a)+h+r$
erasures, insufficient to correct them.
\qed

\begin{example}
\label{ex12}
{\em Consider an $EP(5,2;8,3;3)$ code and
let $d(5,2;8,3;3)$ be its minimum distance.
According to~(\ref{dvhg}), we have 
$d(5,2;8,3;3)\leq \min\{d(2,3,3;a)\,:\,2\leq a\leq 4\}$,
where, by~(\ref{dvhga0}) and~(\ref{dvhgar}),
$d(2,3,3;2)=20$, $d(2,3,3;3)=22$ and $d(2,3,3;4)=21$,
so $d(5,2;8,3;3)\leq 20$.
}
\end{example}

\begin{example}
\label{ex18}
{\em Consider $EP(m,1;n,1;g)$ codes
such that $g+1<\min\{m,n\}$. The following table gives the upper
bound, according to Theorem~\ref{lemma2}, for $d(m,1;n,1;g)$, where
$0\leq g\leq 13$:

\begin{center}
\begin{tabular}{cc}
\begin{tabular}{|c|c|}
\hline
$g$&$d(m,1;n,1;g)\leq $\\
\hline
0&4\\
1&6\\
2&8\\
3&9\\
4&11\\
5&12\\
6&14\\
\hline
\end{tabular}
&
\begin{tabular}{|c|c|}
\hline
$g$&$d(m,1;n,1;g)\leq $\\
\hline
7&15\\
8&16\\
9&18\\
10&19\\
11&20\\
12&22\\
13&23\\
\hline
\end{tabular}
\end{tabular}
\end{center}

}
\end{example}

The special case $v= 0$ in Theorem~\ref{lemma2} corresponds to LRC
codes~\cite{ghsy,tb,pklk}, i.e., there is no vertical code. Let us
state explicitly the result for this case.

\begin{cor}
\label{cor12}
{\em
Consider an $EP(m,0;n,h;g)$ code.
Then,

\begin{eqnarray}
\label{d0hg}
d(m,0;n,h;g)&\hspace{-2mm}\leq\hspace{-2mm} &\left\lceil {\frac {g+1}{n-h}} \right\rceil h\,+\,g\,+\,1
\end{eqnarray}
}
\end{cor}

\noindent\pf
Taking $a=\min\{g+1\,,\,n-h\}$ in~(\ref{dvhg}) gives for this case

\begin{eqnarray}
\label{d0hgmin}
\hspace{-4mm}d(m,0;n,h;g)&\hspace{-2mm}\leq\hspace{-2mm} &d(0,h,g\,;\,\min\{g+1\,,\,n-h\})
\end{eqnarray}

If $g+1<n-h$, then $b= 1$, so~(\ref{dvhga0}) gives
$d(0,h,g\,;\,g+1)= h+g+1$ and~(\ref{d0hg}) follows from~(\ref{d0hgmin}).

If $g+1\,\geq\,n-h$ and $n-h$ divides $g+1$, by~(\ref{dvhga0}),
\begin{eqnarray*}
d(0,h,g;n-h)&\hspace{-2mm}=\hspace{-2mm}&\left({\frac
{g+1}{n-h}}\right)(h+n-h)
\;=\;\left({\frac {g+1}{ n-h}}\right)\hspace{-1mm}h+g+1,
\end{eqnarray*} 
and~(\ref{d0hg}) follows from~(\ref{d0hgmin}). If $g+1>n-h$ and $n-h$
does not divide $g+1$, by~(\ref{dvhgar}),
\begin{eqnarray*}
d(0,h,g;n-h)&\hspace{-2mm}=\hspace{-2mm}&\left\lfloor {\frac
{g\hspace{-.5mm}+\hspace{-.5mm}1}{ n\hspace{-.5mm}-\hspace{-.5mm}h}}
\right\rfloor\hspace{-1mm}
n\hspace{-.5mm}+\hspace{-.5mm}h\hspace{-.5mm}+\hspace{-.5mm}g\hspace{-.5mm}+\hspace{-.5mm}1 
\hspace{-.5mm}-\hspace{-.5mm}(n\hspace{-.5mm}-\hspace{-.5mm}h)\hspace{-1mm}\left\lfloor \hspace{-.5mm}
{\frac {g\hspace{-.5mm}+\hspace{-.5mm}1}{ n\hspace{-.5mm}-\hspace{-.5mm}h}}
\right\rfloor
\;=\;\left(\left\lfloor {\frac {g+1}{ n-h}} \right\rfloor
+1\right)h+g+1
\;=\;\left\lceil {\frac {g+1}{ n-h}} \right\rceil h+g+1,\end{eqnarray*}
and also in this case, (\ref{d0hg}) follows from~(\ref{d0hgmin}).
\qed

Bound~(\ref{d0hg}) is well known, albeit it is usually given in a slightly different
form~\cite{ghsy,pklk} as a function of the dimension of the code,
while bound~(\ref{d0hg}) is given as a function of the redundancy. It
was also shown that bound~(\ref{d0hg}) can be achieved with efficient
constructions~\cite{tb} over a field $GF(q)$, where $q\geq mn$ and in
general $q$ is minimal with this property.

\subsection{Some Optimal Extended Product Codes}
\label{optimalEPC}
We say that an $EP(m,v;n,h;g)$ code is {\em optimal} if it meets
bound~(\ref{dvhg}) with equality.
We believe that there are optimal $EP(m,v;n,h;g)$ codes for any
choice of parameters, but the subject requires further research.

The next theorem shows that there is a range of parameters for which
2-level EII codes are optimal extended product codes.

\begin{theorem}
\label{theo5}
{\rm
Consider the 2-level EII code $\C(n,\uu)$ as given by Definition~\ref{defGPMDS}, where
\begin{eqnarray}
\label{eq3}
\hspace{-5mm}\uu&=&\left(\overbrace{h,h,\ldots,h}^{m-v-1}\,\,,\,h+g\,,\,\,\overbrace{n,n,\ldots,n}^{v}\right),
\end{eqnarray}
$0\leq v<m-1$, $v\leq h$, $h+g<n$ and $1\leq g\leq \left\lceil {\frac{h-v+1}{v+1}}\right\rceil$.
Then, $\C(n,\uu)$ is an optimal $EP(m,v;n,h;g)$ code with minimum
distance
\begin{eqnarray}
\label{eq2}
d&\eq &(h+g+1)(v+1)
\end{eqnarray}
}
\end{theorem}

\noindent\pf
Applying Theorem~\ref{cor11} to~(\ref{eq3}), the minimum distance of $\C(n,\uu)$
is $d\eq\min\left\{(h+g+1)(v+1)\;,\;(h+1)(v+2)\right\}$.
In fact, we will show that
$(h\hspace{-.5mm}+\hspace{-.5mm}g\hspace{-.5mm}+\hspace{-.5mm}1)
(v\hspace{-.5mm}+\hspace{-.5mm}1)\leq (h\hspace{-.5mm}+\hspace{-.5mm}1)(v\hspace{-.5mm}+\hspace{-.5mm}2)$ and
hence~(\ref{eq2}) follows.

In effect, $(h+g+1)(v+1)\leq (h+1)(v+2)$ if and
only if $g\leq (h+1)/(v+1)$. By the conditions on $g$, it suffices to
prove that

\begin{eqnarray}
\label{eq1}
\left\lceil {\frac{h-v+1}{v+1}}\right\rceil&\leq &{\frac{h+1}{v+1}}.
\end{eqnarray}

If $\left\lceil {\frac{h-v+1}{v+1}}\right\rceil\eq {\frac{h-v+1}{v+1}}$,
(\ref{eq1}) is immediate, so assume that
$\left\lceil {\frac{h-v+1}{v+1}}\right\rceil> {\frac{h-v+1}{v+1}}$. Then,
there is a $j$, $1\leq j\leq v$, such that
$\left\lceil {\frac{h-v+1}{v+1}}\right\rceil\eq {\frac{h-v+j+1}{v+1}}$. But
$\frac{h-v+j+1}{v+1}\leq {\frac{h+1}{v+1}}$ since $j\leq v$,
so~(\ref{eq1}) follows also in this case.

Next we have to prove that this minimum distance 
$d$ given by~(\ref{eq2})
meets bound~(\ref{dvhg}) of Theorem~\ref{lemma2}. It suffices to
observe that $d=d(v,h,g;g+1)$ by~(\ref{dvhga0}). 
\qed

Notice that, in particular, if $v\eq 0$, we have an LRC code, as in
Corollary~\ref{cor12}. In this
case, Theorem~\ref{theo5} asserts that when $1\leq g\leq h+1$, then
$\C(n,\uu)$ is an optimal LRC code with minimum distance $d=h+g+1$.
This result was also observed in~\cite{bh}, Corollary~2.3.

Let us examine now the case of $EP(m,1;n,1;2)$ codes, where $m,n\geq 3$. In this case,
bound~(\ref{dvhg}) gives $d(m,1;n,1;2)\leq 8$.

Consider for example a 2-level EII code
$\C(n,(\overbrace{1,1,\ldots,1}^{m-2},3,n))$ or a 2-level EII code
$\C(n,(\overbrace{1,1,\ldots,1}^{m-3},2,2,n))$.
These are the only cases of EII codes that are $EP(m,1;n,1;2)$ codes.
In both cases, according to Theorem~\ref{cor11}, the minimum
distance is 6, so bound~(\ref{dvhg}) is not met.
We present next an optimal
$EP(m,1;n,1;2)$ code. 
The construction is related to the
PMDS constructions in~\cite{bpsy}. The tradeoff is that the
finite field is larger than the required one for EII codes.

Let $GF(2^b)$ be a finite field and $\al$ an element in $GF(2^b)$
such that $mn\,\leq\,\cO(\al)$ (remember, $\cO(\al)$ denotes the order of $\al$).
Consider the 
parity-check matrix $\cH_2$ 
given by

\begin{eqnarray}
\label{pc0}
\hspace{-2mm}\cH_2&\hspace{-2mm}=\hspace{-2mm}&\left(
\begin{array}{c}
I_m\otimes (\overbrace{1,1,\ldots,1}^n)\\
(\overbrace{1,1,\ldots,1}^m)\otimes I_n\\
\hline
\begin{array}{ccccc}
1&\al&\al^2&\ldots &\al^{mn-1}\\
1&\al^{-1}&\al^{-2}&\ldots &\al^{-(mn-1)}\\
\end{array}
\end{array}
\right),
\end{eqnarray}
where $I_m$ denotes the $m\times m$ identity matrix and $\otimes$ the
Kronecker product~\cite{ms} of two matrices. Notice that the first
$m+n$ rows of $\cH_2$ in~(\ref{pc0})
correspond to the parity-check matrix of the product code with
single parity in rows and columns. We denote the matrix
in~(\ref{pc0}) $\cH_2$ to indicate that two extra parities are added
to the product code.

The following theorem
shows that the code whose parity-check matrix is $\cH_2$ is an
optimal $E(m,1;n,1;2)$ code.

\begin{theorem}
\label{lemma3}
{\em
Consider the $EP(m,1;n,1;2)$ code whose parity-check matrix $\cH_2$ is
given by~(\ref{pc0}), $m,n\geq 3$ and $mn\,\leq\,\cO(\al)$. Then, the
code has minimum distance 8.
}
\end{theorem}

\noindent\pf We have to prove that any 7 erasures can be corrected.

First assume that
there are six erasures in locations $(i_0,j_0)$, $(i_0,j_1)$, $(i_0,j_2)$, $(i_1,j_0)$,
$(i_1,j_1)$ and  $(i_1,j_2)$, where $0\leq i_0<i_1\leq m-1$ and $0\leq j_0<j_1<j_2\leq n-1$
or $(i_0,j_0)$, $(i_0,j_1)$, $(i_1,j_0)$, $(i_1,j_1)$,
$(i_2,j_0)$ and  $(i_2,j_1)$, where $0\leq i_0<i_1<i_2\leq m-1$ and $0\leq j_0<j_1\leq n-1$,
and a seventh erasure in any other location. This seventh erasure
can be corrected using either horizontal or vertical parities, thus,
it is enough to prove that the two situations of six erasures
described above are correctable.

Consider the first case.
It suffices to prove, using the
parity-check matrix as given by~(\ref{pc0}) , that the
$6\times 6$ matrix

$$
\left(
\begin{array}{cccccc}
1&1&1&0&0&0\\
0&0&0&1&1&1\\
0&1&0&0&1&0\\
0&0&1&0&0&1\\
\hspace{-1mm}\al^{i_0n+j_0}\hspace{-1mm}&\hspace{-1mm}\al^{i_0n+j_1}\hspace{-1mm}
&\hspace{-1mm}\al^{i_0n+j_2}\hspace{-1mm}&\hspace{-1mm}\al^{i_1n+j_0}\hspace{-1mm}
&\hspace{-1mm}\al^{i_1n+j_1}\hspace{-1mm}&\hspace{-1mm}\al^{i_1n+j_2}\hspace{-1mm}\\
\hspace{-1mm}\al^{-i_0n-j_0}\hspace{-1mm}&\hspace{-1mm}\al^{-i_0n-j_1}\hspace{-1mm}
&\hspace{-1mm}\al^{-i_0n-j_2}\hspace{-1mm}&\hspace{-1mm}\al^{-i_1n-j_0}\hspace{-1mm}
&\hspace{-1mm}\al^{-i_1n-j_1}\hspace{-1mm}&\hspace{-1mm}\al^{-i_1n-j_2}\hspace{-1mm}\\
\end{array}
\right)
$$
is invertible. Redefining $i\la i_1-i_0$, $j_1\la j_1-j_0$ and $j_2\la j_2-j_0$, where
now $1\leq i\leq m-1$ and $1\leq j_1<j_2\leq n-1$, this
matrix is invertible if and only if matrix

$$
\left(
\begin{array}{cccccc}
1&1&1&0&0&0\\
0&0&0&1&1&1\\
0&1&0&0&1&0\\
0&0&1&0&0&1\\
1&\al^{j_1}&\al^{j_2}&\al^{in}&\al^{in+j_1}&\al^{in+j_2}\\
1&\al^{-j_1}&\al^{-j_2}&\al^{-in}&\al^{-in-j_1}&\al^{-in-j_2}\\
\end{array}
\right)
$$
is invertible. By Gaussian elimination, this $6\times 6$ matrix is
invertible if and only if
 the $2\times 2$ matrix
\begin{eqnarray*}
\left(
\begin{array}{ccc}
(1\xor\al^{j_1})(1\xor\al^{in})&(1\xor\al^{j_2})(1\xor\al^{in})\\
(1\xor\al^{-j_1})(1\xor\al^{-in})&(1\xor\al^{-j_2})(1\xor\al^{-in})\\
\end{array}
\right)
\end{eqnarray*}
is invertible, if and only if, since $1\xor\al^{in}\neq 0$,
\begin{eqnarray*}
\left(\hspace{-1mm}
\begin{array}{ccc}
\hspace{-1mm}1\xor\al^{j_1}\hspace{-1mm}&\hspace{-1mm}1\xor\al^{j_2}\hspace{-1mm}\\
\hspace{-1mm}1\xor\al^{-j_1}\hspace{-1mm}&\hspace{-1mm}1\xor\al^{-j_2}\hspace{-1mm}\\
\end{array}
\hspace{-1mm}\right)&\hspace{-1mm}=\hspace{-1mm}&
\left(\hspace{-1mm}
\begin{array}{ccc}
1\xor\al^{j_1}&1\xor\al^{j_2}\\
\hspace{-1mm}\al^{-j_1}(1\xor\al^{j_1})\hspace{-1mm}&\hspace{-1mm}\al^{-j_2}(1\xor\al^{j_2})\hspace{-1mm}\\
\end{array}
\hspace{-1mm}\right)
\end{eqnarray*}
is invertible, if and only if, since $1\hspace{-.5mm}\xor\hspace{-.5mm}\al^{j_1}\hspace{-.5mm}\neq\hspace{-.5mm} 0$ and
$1\hspace{-.5mm}\xor\hspace{-.5mm}\al^{j_2}\hspace{-.5mm}\neq\hspace{-.5mm} 0$,
$\al^{j_1}\hspace{-.5mm}\neq\hspace{-.5mm} \al^{j_2}$, which is the
case since $1\hspace{-.5mm}\leq\hspace{-.5mm} j_1\hspace{-.5mm}
<\hspace{-.5mm}j_2\hspace{-.5mm}\leq\hspace{-.5mm} n-1\hspace{-.5mm}<\hspace{-.5mm}\cO(\al)$.
The second case is proven similarly.

Next, assume that there are seven erasures, such that each row and
column has at least two erasures. This can only happen if one row
(column) has three erasures and two rows (columns) have two erasures.

Let $i_0$ be the row with three erasures, and $j_0$ the column with
three erasures, while $j_1<j_2$ and $i_1$ is such that erasures are
in $(i_1,j_0)$ and $(i_1,j_1)$ so the remaining two erasures are in
$(i_2,j_0)$ and $(i_2,j_2)$.
Redefining $i_1\la i_1-i_0$, $i_2\la i_2-i_0$, $j_1\la j_1-j_0$ and $j_2\la j_2-j_0$,
it suffices to prove, using the
parity-check matrix $\cH_2$ as given by~(\ref{pc0}), that the
$7\times 7$ matrix

$$
\left(
\begin{array}{ccccccc}
1&1&1&0&0&0&0\\
0&0&0&1&1&0&0\\
0&0&0&0&0&1&1\\
0&1&0&0&1&0&0\\
0&0&1&0&0&0&1\\
1&\al^{j_1}&\al^{j_2}&\al^{i_1n}&\al^{i_1n+j_1}&\al^{i_2n}&\al^{i_2n+j_2}\\
1&\al^{-j_1}&\al^{-j_2}&\al^{-i_1n}&\al^{-i_1n-j_1}&\al^{-i_2n}&\al^{-i_2n-j_2}\\
\end{array}
\right)
$$
is invertible, if and only if, doing Gaussian elimination like in the
other two cases, the $2\times 2$ matrix

\begin{eqnarray*}
\left(
\begin{array}{cc}
(1\xor\al^{j_1})(1\xor\al^{i_1n})&(1\xor\al^{j_2})(1\xor\al^{i_2n})\\
(1\xor\al^{-j_1})(1\xor\al^{-i_1n})&(1\xor\al^{-j_2})(1\xor\al^{-i_2n})\\
\end{array}
\right)&=&
\left(\hspace{-2mm}
\begin{array}{cc}
(1\xor\al^{j_1})(1\xor\al^{i_1n})&(1\xor\al^{j_2})(1\xor\al^{i_2n})\\
\al^{-i_1n-j_1}(1\xor\al^{j_1})(1\xor\al^{i_1n})&\al^{-i_2n-j_2}(1\xor\al^{j_2})(1\xor\al^{i_2n})\\
\end{array}
\hspace{-2mm}\right)
\end{eqnarray*}
is invertible, if and only if, since $1\hspace{-.5mm}\xor\hspace{-.5mm}\al^{j_1}$,
$1\hspace{-.5mm}\xor\hspace{-.5mm}\al^{i_1n}$,
$1\hspace{-.5mm}\xor\hspace{-.5mm}\al^{j_2}$  and $1\hspace{-.5mm}\xor\hspace{-.5mm}\al^{i_2n}$ are non-zero,
$\al^{i_1n+j_1}\hspace{-.5mm}\neq\hspace{-.5mm} \al^{i_2n+j_2}$
which is the case
since $mn\,\leq\,\cO(\al)$, thus $(i_2-i_1)n+j_2-j_1\not\equiv 0\;(\bmod\;\cO(\al)).$
For complete details, see~\cite{bh2}.
\qed

Consider next the 3-level EII code
$\C(n,\uu)$, where
$\uu=\left(\overbrace{1,1,\ldots,1}^{m-3},2,3,n\right)$.
This is an
$EP(m,1;n,1;3)$ code. According to Theorem~\ref{cor11},
$\C(n,\uu)$ has minimum
distance 8, the same as the code given by parity-check matrix $\cH_2$,
at the cost of an extra parity. However, there is a tradeoff: the
size of the field required by $\C(n,\uu)$ is
greater than $\max\{m;n\}$, while the field required by the code
whose parity-check matrix is $\cH_2$ must have size greater than
$mn$. Also, by Theorem~\ref{theo2}, $\C(n,\uu)$ can
correct any 8 erasures involving two rows with 3 erasures and one row
with two erasures, like for example
a pattern with erasures in locations $(i_0,j_0)$, $(i_0,j_1)$, $(i_0,j_2)$,
$(i_1,j_0)$, $(i_1,j_1)$, $(i_1,j_2)$, $(i_2,j_0)$ and $(i_2,j_1)$.
The code generated by $\cH_2$ is unable to correct such pattern since
it does not have enough parities; so,
even if both codes have the same minimum distance,
$\C(n,\uu)$ can correct more
erasure patterns. These tradeoffs need to be evaluated when
implementation is considered.

We end this section with a construction of
$EP(m,1;n,1;g)$ codes.
Let $f(x)$ be a binary irreducible polynomial of degree~$b$,
$GF(2^b)$ the field of polynomials modulo $f(x)$ and $\al$ an
element in $GF(2^b)$ such that
$f(\al)\eq 0$.
Consider the following parity-check matrix of a code consisting of
$m\times n$ arrays:

\begin{eqnarray}
\label{pcg}
\hspace{-4mm}\cH(m,n;g)&\hspace{-2mm}=\hspace{-2mm}&\left(
\begin{array}{c}
I_m\otimes (\overbrace{1,1,\ldots,1}^n)\\
(\overbrace{1,1,\ldots,1}^m)\otimes I_n\\
\hline
\begin{array}{ccccc}
1&\al&\al^2&\ldots &\al^{mn-1}\\
1&\al^{2}&\al^{4}&\ldots &\al^{2(mn-1)}\\
\vdots &\vdots &\vdots &\ddots &\vdots \\
1&\al^{g}&\al^{2g}&\ldots &\al^{g(mn-1)}\\
\end{array}
\end{array}
\right)
\end{eqnarray}
Notice that the first $m+n$ rows of $\cH(m,n;g)$ correspond to the parity-check
matrix of a product code with single horizontal and vertical parities, while
the last $g$ rows correspond to the parity-check matrix of an
$[mn,mn-g,g+1]$ RS code over $GF(2^b)$. Assume that
$b\geq \sum_{i=0}^{g-1}\,(i+1)(mn-g+i)$. We will show that the code
$\C(m,n;g)$ whose
parity-check matrix is $\cH(m,n;g)$ as given by~(\ref{pcg}) is an
optimal $E(m,1;n,1;g)$ code. Before proving this result, we need the
following lemma:

\begin{lemma}
\label{lemma4}
{\em
Let
\begin{eqnarray}
\hspace{-4mm}\label{deltag}
\Delta_{j_0,j_1,\ldots,j_{g-1}}(x)&\hspace{-2.5mm}=\hspace{-2.5mm}&\left(
\begin{array}{ccccc}
\hspace{-1mm}x^{j_0}\hspace{-1mm}&\hspace{-1mm}x^{j_1}\hspace{-1mm}&\hspace{-1mm}x^{j_2}\hspace{-1mm}&\ldots &\hspace{-1mm}x^{j_{g-1}}\hspace{-1mm}\\
x^{2j_0}&x^{2j_1}&x^{2j_2}&\ldots &x^{2j_{g-1}}\\
x^{3j_0}&x^{3j_1}&x^{3j_2}&\ldots &x^{3j_{g-1}}\\
\vdots &\vdots &\vdots &\ddots &\vdots \\
x^{gj_0}&x^{gj_1}&x^{gj_2}&\ldots &x^{gj_{g-1}}\\
\end{array}
\right)
\end{eqnarray}
be a $g\times g$ matrix of powers of $x$,
where the $j_i$s are integers such that $0\leq j_0<j_1<\cdots <j_{g-1}$.
Then the binary polynomial $\det\left(\Delta_{j_0,j_1,\ldots,j_{g-1}}(x)\right)$
has degree $\sum_{i=0}^{g-1}(i+1)j_i$.
}
\end{lemma}

\noindent\pf By properties of Vandermonde determinants, $$\det\left(\Delta_{j_0,j_1,\ldots,j_{g-1}}(x)\right)\eq
\left(x^{\sum_{i=0}^{g-1}j_i}\right)\prod_{0\leq u<v\leq {g-1}}(x^{j_u}\xor x^{j_v}).$$
The degree of this polynomial is $\sum_{i=0}^{g-1}(i+1)j_i$ by induction on $g$.
\qed

\begin{theorem}
\label{theo6}
{\rm
Consider the code $\C(m,n;g)$ whose
parity-check matrix is $\cH(m,n;g)$ as given by~(\ref{pcg}), where
$GF(2^b)$ is the field of polynomials modulo the binary irreducible polynomial $f(x)$,
$f(\al)\eq 0$ and
$b\geq \sum_{i=0}^{g-1}\,(i+1)(mn-g+i)$. Then, $\C(m,n;g)$ is
an optimal $E(m,1;n,1;g)$ code.
}
\end{theorem}

\noindent\pf Assume that
$d$ is the upper bound on the minimum distance of a code $E(m,1;n,1;g)$
given by Theorem~\ref{lemma2}.
We will prove that any $d-1$ erasures can be
corrected by code $\C(m,n;g)$. So, assume that we have $d-1$
erasures, say, in locations
$0\leq j_0<j_1<\cdots <j_{d-2}\leq mn-1$.
From
Theorem~\ref{lemma2}, there are $d-1-g$ horizontal and vertical
parities covering the erasures that are linearly independent
(otherwise, we would be violating the bound).
Consider the $(d-1)\times
(d-1)$ sub-matrix of $\cH(m,n;g)$ whose entries are given by the
intersection of columns $j_0,j_1,\ldots,j_{d-2}$
with the
rows corresponding to the aforementioned $d-1-g$
linearly independent horizontal and vertical parities, followed by
the last $g$ rows of $\cH(m,n;g)$. We have to prove that this matrix
is invertible. The matrix looks as follows:

\begin{eqnarray*}
H_{d-1}&=&\left(
\begin{array}{c}
V\\
\hline
W
\end{array}
\right),
\end{eqnarray*}
where $V$ is a $(d-1-g)\times (d-1)$ matrix of rank $d-1-g$ whose
entries are 0s and 1s and

\begin{eqnarray*}
W&=&\left(\begin{array}{cccc}
\al^{j_0}&\al^{j_1}&\ldots &\al^{j_{d-2}}\\
\al^{2j_0}&\al^{2j_1}&\ldots &\al^{2j_{d-2}}\\
\vdots &\vdots &\ddots &\vdots \\
\al^{gj_0}&\al^{gj_1}&\ldots &\al^{gj_{d-2}}\\
\end{array}\right).
\end{eqnarray*}

If 
$U\subseteq \{0,1,\ldots,d-2\}$
and $\overline{U}\eq
\{0,1,\ldots,d-2\}-U$, denote by $V[U]$ the columns of $V$ in
locations $U$ and by $W[\overline{U}]$ the columns of $W$ in $\overline{U}$.
By properties of determinants and since $V$ is a binary matrix,

\begin{eqnarray}
\label{Hdm1}
\det H_{d-1}&=&\bigoplus_{\substack{U\subseteq 
\{0,1,\ldots,d-2\}\\|U|=d-1-g}}
(\det V[U])(\det W[\overline{U}])
\;=\;\bigoplus_{\substack{U\subseteq \{0,1,\ldots,d-2\}\\|U|=d-1-g\,,\,\det V[U]=1}}
\det W[\overline{U}].
\end{eqnarray}

Since $V$ has rank $d-1-g$, let $U_0$ be the first subset of $\{0,1,\ldots,d-2\}$
in alphabetical order such
that $|U_0|\eq d-1-g$ and $\det V[U_0]=1$. Then, $\overline{U}_0$ is the last subset in
alphabetical order such that $\det V[U_0]=1$, i.e., if $U\neq U_0$,
$\det V[U]=1$,  $\overline{U}\eq\{u_0,u_1,\ldots,u_{g-1}\}$ and
$\overline{U}_0\eq\{w_0,w_1,\ldots,w_{g-1}\}$, where $u_0<u_1<\cdots <u_{g-1}$
and $w_0<w_1<\cdots <w_{g-1}$,
then $u_i\leq w_i$ for
$0\leq i\leq g-1$. By Lemma~\ref{lemma4},
$\det W[\overline{U}]$ has degree (as a polynomial in~$\al$)
$\sum_{i=0}^{g-1}\,(i+1)j_{u_i}\,<\,\sum_{i=0}^{g-1}\,(i+1)j_{w_i}$, which is the degree of
$\det W[\overline{U}_0]$. This means, $\al^{\sum_{i=0}^{g-1}\,(i+1)j_{w_i}}$ cannot be canceled
by any other power of $\al$ in~(\ref{Hdm1}) and $\det H_{d-1}$ is a
polynomial in $\al$ of degree $\sum_{i=0}^{g-1}\,(i+1)j_{w_i}$.
Since
$\sum_{i=0}^{g-1}\,(i+1)j_{w_i}\,<\,\sum_{i=0}^{g-1}\,(i+1)(mn-g+i)\,\leq b$,
$\det H_{d-1}\neq 0$ and $H_{d-1}$ is invertible.
\qed

We illustrate the proof of Theorem~\ref{theo6} in the following example:

\begin{example}
\label{ex19}
{\em
Consider the code $\cC(4,9;3)$ whose parity-check matrix is
$\cH(4,6;3)$ as given by~(\ref{pcg}), and
assume that $b\geq 33+(2)(34)+(3)(35)\eq 206$. For instance, we
may take $\cC(4,9;3)$ over the field $GF(2^{206})$
with $f(x)$ an irreducible polynomial of degree 206 and $f(\al)\eq 0$.
According to Theorems~\ref{lemma2}
and~\ref{theo6}, the minimum distance of this code is $d\eq 9$, i.e.,
any 8 erasures can be corrected. In effect, assume that we have the
following array with 8 erasures:

$$
\begin{array}{|c|c|c|c|c|c|c|c|c|}
\hline
\phantom{E}&E&\phantom{E}&\phantom{E}&E&\phantom{E}&\phantom{E}&\phantom{E}&\phantom{E}\\
\hline
\phantom{E}&\phantom{E}&\phantom{E}&\phantom{E}&\phantom{E}&\phantom{E}&\phantom{E}&\phantom{E}&\phantom{E}\\
\hline
E&E&\phantom{E}&\phantom{E}&E&E&\phantom{E}&\phantom{E}&\phantom{E}\\
\hline
E&\phantom{E}&\phantom{E}&\phantom{E}&\phantom{E}&E&\phantom{E}&\phantom{E}&\phantom{E}\\
\hline
\end{array}
$$

Following the proof of Theorem~\ref{theo6}, erasures have occurred in
locations
$$\{j_0,j_1,j_2,j_3,j_4,j_5,j_6,j_7\}\eq\{1,4,18,19,22,23,27,32\}.$$
Using the parity-check matrix $\cH(4,9;3)$ given by~(\ref{pcg}),
it suffices to prove that
the following $8\times 8$ determinant in $\al$ is non-zero:

\begin{eqnarray}
\label{galpha}
\hspace{-3mm}g(\al)&\hspace{-2mm}=\hspace{-2mm}&
\det\hspace{-1mm}\left(\hspace{-1mm}
\begin{array}{cc|cccc|cc}
1&1&0&0&0&0&0&0\\
0&0&1&1&1&1&0&0\\
0&0&0&0&0&0&1&1\\
\hline
1&0&0&1&0&0&0&0\\
0&0&1&0&0&0&1&0\\
\hline
\hspace{-1mm}\al\hspace{-1mm} &\hspace{-1mm}\al^4\hspace{-1mm}&
\hspace{-1mm}\al^{18}\hspace{-1mm}&\hspace{-1mm}\al^{19}\hspace{-1mm}&\hspace{-1mm}\al^{22}\hspace{-1mm}
&\hspace{-1mm}\al^{23}\hspace{-1mm}&\hspace{-1mm}\al^{27}\hspace{-1mm}&\hspace{-1mm}\al^{32}\hspace{-1mm}\\
\hspace{-1mm}\al^2\hspace{-1mm} &\hspace{-1mm}\al^8\hspace{-1mm}&
\hspace{-1mm}\al^{36}\hspace{-1mm}&\hspace{-1mm}\al^{38}\hspace{-1mm}&\hspace{-1mm}\al^{44}\hspace{-1mm}
&\hspace{-1mm}\al^{46}\hspace{-1mm}&\hspace{-1mm}\al^{54}\hspace{-1mm}&\hspace{-1mm}\al^{64}\hspace{-1mm}\\
\hspace{-1mm}\al^3\hspace{-1mm} &\hspace{-1mm}\al^{12}\hspace{-1mm}&
\hspace{-1mm}\al^{54}\hspace{-1mm}&\hspace{-1mm}\al^{57}\hspace{-1mm}&
\hspace{-1mm}\al^{66}\hspace{-1mm}&\hspace{-1mm}\al^{69}\hspace{-1mm}&\hspace{-1mm}\al^{81}\hspace{-1mm}&
\hspace{-1mm}\al^{96}\hspace{-1mm}\\
\end{array}
\hspace{-1mm}\right)
\end{eqnarray}

The first invertible $5\times 5$ submatrix of the first 5 rows is

$$
\left(
\begin{array}{ccccc}
1&1&0&0&0\\
0&0&1&1&0\\
0&0&0&0&1\\
\hline
1&0&0&1&0\\
0&0&1&0&1\\
\end{array}
\right),
$$
which corresponds to the subset of columns $U_0\eq\{0,1,2,3,6\}$. The
complement of this set of columns is $\overline{U}_0\eq\{4,5,7\}$,
so, since $\{j_4,j_5,j_7\}\eq\{22,23,32\}$, the degree of $g(\al)$
in~(\ref{galpha}) corresponds to the degree of
$\Delta_{22,23,32}(\al)$ which, by Lemma~\ref{lemma4}, is
$22+(2)(23)+(3)(32)\eq 164$. Since $b\eq 206$, $g(\al)$ is
non-zero.
}
\end{example}

Theorem~\ref{theo6} provides an infinite family of $E(m,1;n,1;g)$
codes. It is sufficient to use a code $\cC(m,n;g)$ over a field
$GF(2^b)$ with $b\geq\sum_{i=0}^{g-1}\,(i+1)(mn-g+i)$.
However, 
from a practical point of view, this process
requires a very large finite field with the
corresponding increase in complexity even for relatively small values of
$m$, $n$ and $g$, as we have seen in Example~\ref{ex19}.

A way to overcome this problem and make the codes practical for
implementation is to use
the field $GF(2^{p-1})$, $p$ a prime number, such that $GF(2^{p-1})$
is generated by the irreducible polynomial $M_p(x)=1+x+x^2+\cdots
+x^{p-1}$. This field was often used in array codes requiring symbols
of large size~\cite{br}.
The polynomial $M_p(x)$ is not irreducible for
every prime number $p$. For
example, $M_5(x)$ is irreducible but $M_7(x)=
(1+x+x^3)(1+x^2+x^3)$. For an irreducible $M_p(x)$, if $M_p(\al)=0$, then $\al^p=1$. If we
choose $M_p(x)$ and $p$ is a prime number large enough, then we can apply 
Theorem~\ref{theo6} and
the code will be optimal. 
We state this result as a corollary.

\begin{cor}
\label{cor5}
{\em
Consider the $EP(m,1;n,1;g)$ code whose parity-check matrix is given
by~(\ref{pcg}) with $\al$ in~(\ref{pcg}) a zero of $M_p(x)$, $p$ a
prime number, $M_p(x)$ irreducible and
$\sum_{i=0}^{g-1}\,(i+1)(mn-g+i)\leq p-1$. Then the code is an optimal
$EP(m,1;n,1;g)$ code.
}
\end{cor}



Although the field of polynomials modulo
$M_p(x)$ has size $2^{p-1}$ (possibly a very large number), no look-up tables are necessary in
implementation, since most operations reduce to XORs and
rotations~\cite{br}, but we omit the details here.
Strictly speaking, it
is not proven that the number of primes such that $M_p(x)$ is
irreducible is infinite, 
but from a practical point of view, it is always
possible to find such a large enough prime number.

\subsection{Some Performance Considerations}
\label{performance}
It is known that in product codes, the row-column iterative decoding algorithm
does not necessarily correct all the correctable erasure
patterns~\cite{ghswy,jbs}. Similarly, there may be correctable erasure patterns in a
$t$-level EII code that cannot be corrected by the row-column iterative decoding
algorithm. In effect, if after applying the iterative decoding
algorithm there are still erasures left, sometimes such erasures may be
corrected by solving a
linear system using the parity-check matrix of the code. However, we
do not deal here with this residual erasure correcting capability. A
natural question is, how much does the row-column iterative decoding
algorithm enhance the individual row or column decoding algorithms?
The answer depends on the particular parameters considered. For
example, take a 4-level II code $\C(7,(1,2,3,6,6))$. By
Theorem~\ref{cor11}, the minimum distance of this code is $d\eq 7$.
Assume that the erasures, which may correspond to failures of whole
storage devices, occur randomly, one after the other.
An important parameter is the average number of erasures that will
produce an uncorrectable pattern~\cite{bh}. If we decode by rows only, a
Monte Carlo simulation for this example gives that this average number is 14.1. The
column code, by Theorem~\ref{theo3}, is a 4-level EII code
$\C(5,(0,2,2,2,3,4,5))$. Decoding by columns, the
Monte Carlo simulation gives that the average number of erasures
producing an uncorrectable pattern is 13.3. The iterative row-column
decoding algorithm gives an average of 15.3, better than the other
two algorithms taken separately. Another way of looking at the
performance of the three algorithms is as follows: assume that a (random)
number of erasures greater than the minimum distance has occurred.
What is the probability that each of the algorithms will correct such
pattern? For example, taking the same code $\C(7,(1,2,3,6,6))$,
assume that 13 erasures have occurred. Again by Monte Carlo simulation,
we found out that the row decoding algorithm corrects 64\% of such
patterns, the column decoding algorithm corrects 49\%, while the
iterative row-column decoding algorithm corrects 84\% of them.

The average number of
erasures causing an uncorrectable erasure pattern in a $t$-level EII
code is very related to 
the mean time to data loss (MTTDL)~\cite{g,gm,gm2} in RAID type of
architectures, specially when failures occur following a Poisson
model~\cite{gm2}, and  
to birthday surprise type of
problems~\cite{kn}. For example, assume that we have a RAID~5 type of architecture, where
each row of an $m\times n$ array is in an $[n,n-1,2]$ code. Using the
notation of $t$-level EII codes, this scheme corresponds to a 1-level
II code $\C(n,(\overbrace{1,1,\ldots,1}^m))$. When
erasures start occurring, there will be an uncorrectable pattern when
one row has two erasures. What is the average number of erasures
causing this uncorrectable pattern? 
This question is equivalent to the birthday surprise problem: assuming that people
arrive at random in a planet whose year has $m$ days, what is the
average number of people that arrive until two of them have the same birthday?
An exact formula for this number is well known, mainly, it is
$m\int_0^{\infty}e^{-mx}(1+x)^m\,dx$~\cite{kn}. On Earth, $m=365$ and this average gives
24.6, the birthday surprise number. It is possible to obtain  
exact formulae for the average of general $t$-level EII codes, but such
formulae become too complicated. The Monte Carlo simulations
provide good approximations though.

We end this subsection with an example comparing an EII code with other types
of codes. Consider a code of length 64 and rate 1/2. Certainly, an
MDS code has minimum distance 33, but an (extended) RS code requires the code to
be at least over the field $GF(64)$. If we want a smaller field like
$GF(16)$, we can use, for instance, Algebraic Geometry (AG) codes.
In~\cite{hlp}, page 23, a $[64,32,27]$ AG code over $GF(16)$ is
presented.  Consider a 5-level II code $\C(8,(2, 3, 3, 4, 4, 5, 5,
6))$, also over $GF(16)$. This code, by Theorem~\ref{cor11}, is a
$[64,32,7]$ code, so its minimum distance is considerably smaller
than the one of the AG code. However, the average number of erasures
that are uncorrectable, by Monte Carlo simulation, is 30.1. If the AG
does not correct erasures beyond its minimum distance, the average
number of uncorrectable erasures is precisely the minimum distance
27, meaning that on average the AG code corrects less erasures than
$\C(8,(2,3,3,4,4,5,5,6))$. Also, the AG code, as well as the RS code,
have no
locality properties. In particular, $\C(8,(2,3,3,4,4,5,5,6))$
is an LRC code of length 64 and locality 6. In addition to the 16
local parities, there are 16 extra (global) parities. By~(\ref{d0hg}),
an upper bound on the minimum distance of such a code
is 23. Assuming a code meeting the bound (i.e., optimal) is used and
there is no correction
beyond the minimum distance once the local erasures have been corrected, 
a Monte Carlo simulation gives that the average number of
uncorrectable erasures is 27, which is below 30.1 as given
by the EII code.
Another way of looking at the problem is the following: assuming that
a number of erasures have occurred, what are the probabilities that an optimal LRC code
or the row-column
decoding algorithm of $\C(8,(2,3,3,4,4,5,5,6))$ will decode such a
pattern? For example, assuming that 27 erasures have occurred, a Monte Carlo simulation
gives that an optimal LRC code can correct roughly 50\% of such
patterns, while $\C(8,(2,3,3,4,4,5,5,6))$ can correct 88\% of them. The
best constructions of optimal LRC codes would require a field
of size at least the length of the code~\cite{tb}, 64 in this case.

The  examples given above show that there are tradeoffs to be considered
when choosing an EII or an optimal LRC code in applications.

\section{Conclusions}
\label{conclusions}
We have studied extended product (EPC) codes,
which consist of a product code with some extra parities added in
order to increase
the minimum distance. We presented an upper bound on the minimum
distance of EPC codes and we gave constructions of codes
achieving this upper bound for the case in which the product code
consists of single parity on rows and columns.
We also studied in detail a special case of EPC codes: Extended
Integrated Interleaved (EII) codes, which in general do not meet the
bound on the minimum distance, but require a small finite field and
allow for a large variety of possible parameters, making them an attractive
alternative for implementation in practical cases. We showed that EII
codes naturally unify product codes and Integrated
Interleaved (II) codes. We provided the distance, the dimension and
encoding and (erasure) decoding algorithms for any EII code. We showed
that EII codes often have better minimum
distance than II codes with the same rate, they allow for
decoding on columns as well as on rows (enhancing the correction capability of
the decoding algorithm) and they permit an
uniform distribution of the parity in the array.

\section*{Acknowledgment}
We want to thank Prof. Yuval Cassuto for pointing out the problem of
balanced distribution of parity in Integrated Interleaved codes.


\begin{thebibliography}{99}






\bibitem{bhh}
M. Blaum, J. L. Hafner and S. Hetzler, ``Partial-MDS Codes and their
Application to RAID Type of Architectures,''
IEEE Trans. on Information Theory, vol. IT-59, pp.~4510–-19,
July 2013.

\bibitem{bh}
M. Blaum and S. Hetzler, ``Integrated Interleaved Codes as Locally
Recoverable Codes,'' Int. J. Information and Coding
Theory, Vol. 3, No. 4, pp.~324--44, September 2016.

\bibitem{bh2}
M. Blaum and S. Hetzler, ``Generalized and Extended Product Codes,''
arXiv:1610.04273, October 2016.
	

\bibitem{bpsy}
M. Blaum, J. S. Plank, M. Schwartz and E. Yaakobi,
``Partial MDS (PMDS) and Sector-Disk (SD) Codes that Tolerate the
Erasure of Two Random Sectors,''
IEEE Trans. on Information Theory, vol. IT-62, pp.~2673--81, May 2016.

\bibitem{br}
M. Blaum and R. M. Roth, ``New Array Codes for Multiple Phased Burst Correction,''
IEEE Trans. on Information Theory, vol. IT-39,
pp.~66-77, January 1993.

\bibitem{bz}
E. L. Blokh and V. V. Zyablov, ``Coding of Generalized Concatenated
Codes,'' Problemy Peredachii Informatsii, Vol. 10(3), pp.~218--222, 1974.



\bibitem{chpb}
Y. Cassuto, E. Hemo, S. Puchinger and M. Bossert,
``Multi-Block Interleaved Codes for Local and Global Read Access,''
ISIT 2017, IEEE International Symposium on
Information Theory, pp.~1758--62, July 2017.

\bibitem{kmc} K. M. Cheung,
``Error-Correction Coding in Data Storage Systems,''
Caltech Ph. D. Thesis, 1987.

\bibitem{kmc2} K. M. Cheung,
``More on the Decoder Error Probability for Reed-Solomon Codes,''
IEEE Trans. on Information Theory, vol. IT-35, pp.~895--900, July 1989.







\bibitem{g}
G. A. Gibson, ``Redundant Disk Arrays,'' MIT Press, 1992.

\bibitem{gm} R. M. F. Goodman and R. J. McEliece,
``Lifetime Analyses of Error-Control
coded semiconductor RAM systems,'' Proc. IEE, part E, vol. 3, pp.
81--85, 1982.

\bibitem{gm2} R. M. F. Goodman and R. J. McEliece,
``Hamming Codes, Computer Memories, and the Birthday
Surprise,'' Proceedings of 20th Allerton Conference on  Communication, Control,
and Computing, pp.~672--79, October 1982.

\bibitem{ghjy} P. Gopalan, C. Huang, B. Jenkins and S. Yekhanin,
``Explicit Maximally Recoverable Codes with Locality,''
IEEE Trans. on Information Theory, vol. IT-60,
pp.~5245--56, September 2014.

\bibitem{ghsy} P. Gopalan, C. Huang, H. Simitci and S. Yekhanin,
``On the Locality of Codeword Symbols,''
IEEE Trans. on Information Theory, vol. IT-58, pp.~6925--34, November
2012.

\bibitem{ghswy} P. Gopalan, G. Hu, S. Saraf, C. Wang and S. Yekhanin,
``Maximally Recoverable Codes for Grid-like Topologies,'' arXiv
1605.05412v1, May 2016.

\bibitem{gl} P. S. Guinand and J. H. Lodge, ``Graph Theoretic
Construction of Generalized Product Codes,''
ISIT 1997, IEEE International Symposium on
Information Theory, p.~111, June 1997.





\bibitem{hp} C. H\"ager, H. D. Pfister, A. G. i Amat and F. Br\"annstr\"om,
``Density Evolution for Deterministic Generalized Product Codes on
the Binary Erasure Channel at High Rates,''
IEEE Trans. on Information Theory, vol. IT-63, pp.~4357--78, July 2017.

\bibitem{hapkt}
M. Hassner, K. Abdel-Ghaffar, A. Patel, R. Koetter and B. Trager,
``Integrated Interleaving -- A Novel ECC Architecture,'' IEEE
Transactions on Magnetics, Vol. 37, No. 2, pp.~773--5, March 2001.

\bibitem{hlp}
T. H\o holdt, J. H. van Lint and R. Pellikaan,
``Algebraic Geometry Codes,''
Handbook of Coding Theory, vol I, pp. 871--961,
(V. S. Pless, W. C. Huffman and R. A. Brualdi Eds.),
Elsevier, Amsterdam, 1998 (corrected version, September 20, 2011).

\bibitem{hy}
G. Hu and S. Yekhanin, ``New Constructions of SD and MR Codes over
Small Finite Fields,'' ISIT 2016, IEEE International Symposium on
Information Theory, pp.~1591--95, July 2016.

\bibitem{hcl}
C. Huang, M. Chen, and J. Li, ``Pyramid Codes: Flexible
Schemes to Trade Space for Access Efficiency in Reliable Data
Storage Systems,'' Proc. of IEEE NCA, Cambridge, Massachussetts, July
2007.

\bibitem{hsx}
C. Huang, H. Simitci, Y. Xu, A. Ogus, B. Calder, P. Gopalan, J. Li
and S. Yekhanin,
``Erasure Coding in Windows Azure Storage,'' 2012 USENIX Annual
Technical Conference, Boston, Massachussetts, June 2012.




\bibitem{jbs}
F. Jardel, J. J. Boutros and M. Sarkiss, ``Stopping Sets for
MDS-Based Product Codes,'' ISIT 2016, IEEE International Symposium on
Information Theory, pp.~1740--44, July 2016.




\bibitem{kn}
M. S. Klamkin and D. J. Newman, ``Extensions of the Birthday
Surprise,'' J. Combin. Theory 3, pp.~279--82, 1967.

\bibitem{kna}
M. Kuijper and D. Napp, ``Erasure Codes with Simplex Locality,''
21st International Symposium on Mathematical Theory of Networks and Systems,
Groningen, The Netherlands, July 2014.



\bibitem{ll}
M. Li and P. C. Lee,
``STAIR Codes: A General Family of Erasure Codes
for Tolerating Device and Sector Failures in
Practical Storage Systems,''
12th USENIX Conference on File and Storage Technologies (FAST '14),
Santa Clara, CA, February 2014.

\bibitem{ld} X. Li and I. Duursma, ``Sector-Disk Codes with Three
Global Parities,'' ISIT 2017, IEEE International Symposium on
Information Theory, pp.~609--13, July 2017.





\bibitem{ms} F. J. MacWilliams and N. J. A. Sloane, ``The Theory
of Error-Correcting Codes,'' North Holland, Amsterdam, 1977.


\bibitem{M}
Micron, ``N-29-17: NAND Flash Design and Use Considerations
Introduction,'' 
http://download.micron.com/pdf/technotes/nand/tn2917.pdf.



\bibitem{msw}
R. J. McEliece and L. Swanson, ``On the Decoder Error Probability for
Reed-Solomon Codes,''
IEEE Trans. on Information Theory, vol. IT-32, pp.~701--03, September 1986.

\bibitem{pb}
J. S. Plank and M. Blaum, ``Sector-Disk (SD) Erasure Codes for Mixed Failure Modes
in RAID Systems,'' ACM Transactions on Storage, Vol. 10, No. 1,
Article 4, January 2014.

\bibitem{pbh}
J. S. Plank, M. Blaum and J. L. Hafner, ''SD Codes:
Erasure Codes Designed for how Storage Systems
Really Fail,'' 11th USENIX
Conference on File and Storage Technologies (FAST '13),
Santa Clara, CA, February 2013.



\bibitem{pd}
D. S. Papailiopoulos and A. G. Dimakis,
``Locally Repairable Codes,'' IEEE Trans. on Information Theory, vol.
IT-60, pp.~5843–-55, October 2014.

\bibitem{pklk}
N. Prakash, G. M. Kamath, V. Lalitha and P. V. Kumar, ``Optimal
Linear Codes with a Local-Error-Correction Property,''
ISIT 2012, IEEE International Symposium on
Information Theory, pp.~2776--80, July 2012.


\bibitem{rk}
A. S. Rawat, O. O. Koyluoglu, N. Silberstein and S. Vishwanath,
``Optimal Locally Repairable and Secure Codes for Distributed Storage Systems,''
IEEE Trans. on Information Theory, vol. IT-60, pp.~212–-36,
January 2014.

\bibitem{rp}
A. S. Rawat, D. S. Papailiopoulos, A. G. Dimakis and S. Vishwanath,
``Locality and Availability in Distributed Storage,'' arxiv:1402.2011,
February 2014.

\bibitem{sa}
M. Sathiamoorthy, M. Asteris, D. Papailiopoulos, A. G. Dimakis,
R. Vadali, S. Chen and D. Borthakur,
''XORing Elephants: Novel Erasure Codes for Big Data,''
Proceedings of VLDB, Vol.~6, No.~5, pp.325--36, 2013.

\bibitem{sd}
W. Song, S. H. Dau, C. Yuen and T. J. Li,
``Optimal Locally Repairable Linear Codes,''
IEEE Journal on Selected Areas in Communications,
Vol.~32 , pp.~1019--36, May 2014.

\bibitem{tb}
I. Tamo and A. Barg,
``A Family of Optimal Locally
Recoverable Codes,'' IEEE Trans. on Information Theory, vol. IT-60,
pp.~4661--76, August 2014.


\bibitem{tk} X. Tang and R. Koetter, ``A Novel Method for Combining
Algebraic Decoding and Iterative Processing,''
ISIT 2006, IEEE International Symposium on Information Theory,
pp.~474--78, July 2006.










\bibitem{wz}
A. Wang and Z. Zhang,
``Repair Locality with Multiple Erasure Tolerance,''
IEEE Trans. on Information Theory, vol. IT-60, pp.~6979--87, November 2014.

\bibitem{w}
Y. Wu, ``Generalized Integrated Interleaved Codes,''
IEEE Trans. on Information Theory, vol. IT-63,
pp.~1102--19, February 2017.

\bibitem{zy}
A. Zeh and E. Yaakobi, ``Bounds and Constructions of Codes with
Multiple Localities,''  ISIT 2016, IEEE International Symposium on
Information Theory, pp.~640--44, July 2016.

\bibitem{z2} X. Zhang, ``Modified Generalized Integrated Interleaved
Codes for Local Erasure Recovery,'' IEEE Communications Letters, Vol.
21, No. 6, pp.~1241--44, June 2017.

\bibitem{z} V. A. Zinoviev, ``Generalized Cascade Codes,'' Problemy
Peredachii Informatsii, vol. 12, no. 1, pp.~5–-15, 1976.

\end{thebibliography}
\end{document}